\documentclass[sigconf]{acmart}

\usepackage{makecell}

\usepackage{tfrupee}
\usepackage{multirow}
\usepackage{subcaption}
\usepackage{enumitem}
\usepackage{soul}
\usepackage{listings}

\usepackage{booktabs}
\usepackage{siunitx}
\usepackage{etoolbox}
\usepackage{framed}
\usepackage{xcolor}
\usepackage{amssymb}
\usepackage{calc}
\usepackage{url}
\usepackage{hyperref}
\AtBeginDocument{%
  }

\setcopyright{acmlicensed}
\copyrightyear{2018}
\acmYear{2018}
\acmDOI{XXXXXXX.XXXXXXX}
\acmConference[Conference acronym 'XX]{Make sure to enter the correct
   conference title from your rights confirmation email}{June 03--05,
   2018}{Woodstock, NY}
\acmISBN{978-1-4503-XXXX-X/2018/06}

\copyrightyear{2026}
\acmYear{2026}
\setcopyright{cc}
\setcctype{by}
\acmConference[CHI '26]{Proceedings of the 2026 CHI Conference on Human Factors in Computing Systems}{April 13--17, 2026}{Barcelona, Spain}
\acmBooktitle{Proceedings of the 2026 CHI Conference on Human Factors in Computing Systems (CHI '26), April 13--17, 2026, Barcelona, Spain}
\acmPrice{}
\acmDOI{10.1145/3772318.3790519}
\acmISBN{979-8-4007-2278-3/2026/04}

\begin{document}

\title{TALES: A Taxonomy and Analysis of Cultural Representations in LLM-generated Stories}

\newcommand{\github}{\raisebox{-1.5pt}{\includegraphics[height=1.05em]{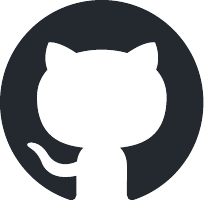}}\xspace}

\definecolor{boxbg}{RGB}{245, 245, 255}   
\definecolor{boxborder}{RGB}{50, 50, 50}   
\setlength{\fboxrule}{0.7pt} 
\setlength{\fboxsep}{5.5pt}
\newenvironment{quotebox}
  {\vspace{3pt} %
   \def\FrameCommand{\fcolorbox{boxborder}{boxbg}}%
   \MakeFramed{\hsize=0.9\linewidth \advance\hsize-\width \FrameRestore}%
   \setlength{\parindent}{0pt}%
   \setlength{\parskip}{0pt}%
   \vspace{3pt} %
  }
  {\vspace{2pt}\endMakeFramed
   \vspace{0pt} %
  }

\newcommand{\munepsfig}[4][scale=1.0]{%
    \begin{figure}[t]
        \centering
        \vspace{2mm}
        \setlength{\fboxrule}{#4} %
        \framebox{\includegraphics[#1]{#2.png}} %
        \caption{#3}
        \label{fig:#2}
    \end{figure}
}

\newlist{questions}{enumerate}{1}
\setlist[questions,1]{label=\textbf{RQ\arabic*:},ref=RQ\arabic*}

\author{Kirti Bhagat}
\authornote{Core contributors, correspondence at: \texttt{\href{mailto:kirtibhagat@iisc.ac.in}{kirtibhagat@iisc.ac.in}} and \texttt{\href{mailto:shaily@cmue.edu}{shaily@cmu.edu}}}
\affiliation{%
 \institution{Indian Institute of Science}
 \city{Bengaluru}
 \state{Karnataka}
 \country{India}}
\email{kirtibhagat@iisc.ac.in}

\author{Shaily Bhatt}
\authornotemark[1]
\affiliation{%
 \institution{Carnegie Mellon University}
 \city{Pittsburgh}
 \state{Pennsylvania}
 \country{USA}}
\email{shaily@cmu.edu}

\author{Athul Velagapudi}
\affiliation{%
\institution{Indian Institute of Science}
 \city{Bengaluru}
 \state{Karnataka}
 \country{India}}
\email{velagapudia@iisc.ac.in}

\author{Aditya Vashistha}
\affiliation{%
 \institution{Cornell University}
 \city{Ithaca}
 \state{New York}
 \country{USA}}
\email{adityav@cornell.edu}

\author{Shachi Dave}
\authornote{Author did not participate in any experimental analysis with LLaMA Models.}
\affiliation{%
 \institution{Google DeepMind}
 \city{Bengaluru}
 \state{Karanataka}
 \country{India}}
\email{shachi@google.com }

\author{Danish Pruthi}
\affiliation{%
\institution{Indian Institute of Science}
 \city{Bengaluru}
 \state{Karnataka}
 \country{India}}
\email{danishp@iisc.ac.in}

\renewcommand{\shortauthors}{Bhagat and Bhatt et al.}

\begin{abstract}  
Millions of users across the globe turn to AI chatbots for their creative needs, inviting widespread interest in understanding how they represent diverse cultures. 
However, evaluating cultural representations in open-ended tasks remains challenging and underexplored. 
In this work, we present TALES, an evaluation of cultural misrepresentations in LLM-generated stories for diverse Indian cultural identities. 
First, we develop TALES-Tax, a taxonomy of cultural misrepresentations by collating insights from participants with lived experiences in India through focus groups (N=$9$) and individual surveys (N=$15$).
Using TALES-Tax, we evaluate $6$ models through a large-scale annotation study spanning $2,925$ annotations from $108$ annotators with lived experience and native language proficiency from across $71$ regions in India and $14$ languages. 
Concerningly, we find that $88$\% of the generated stories contain misrepresentations, and such errors are more prevalent in mid- and low-resourced languages and stories based in peri-urban regions in India. 
We also transform the annotations into TALES-QA, a standalone question bank to evaluate the cultural knowledge of models.
\end{abstract}



\maketitle

\section{Introduction}

\begin{figure*}[t]
  \centering
  \includegraphics[width=\textwidth]{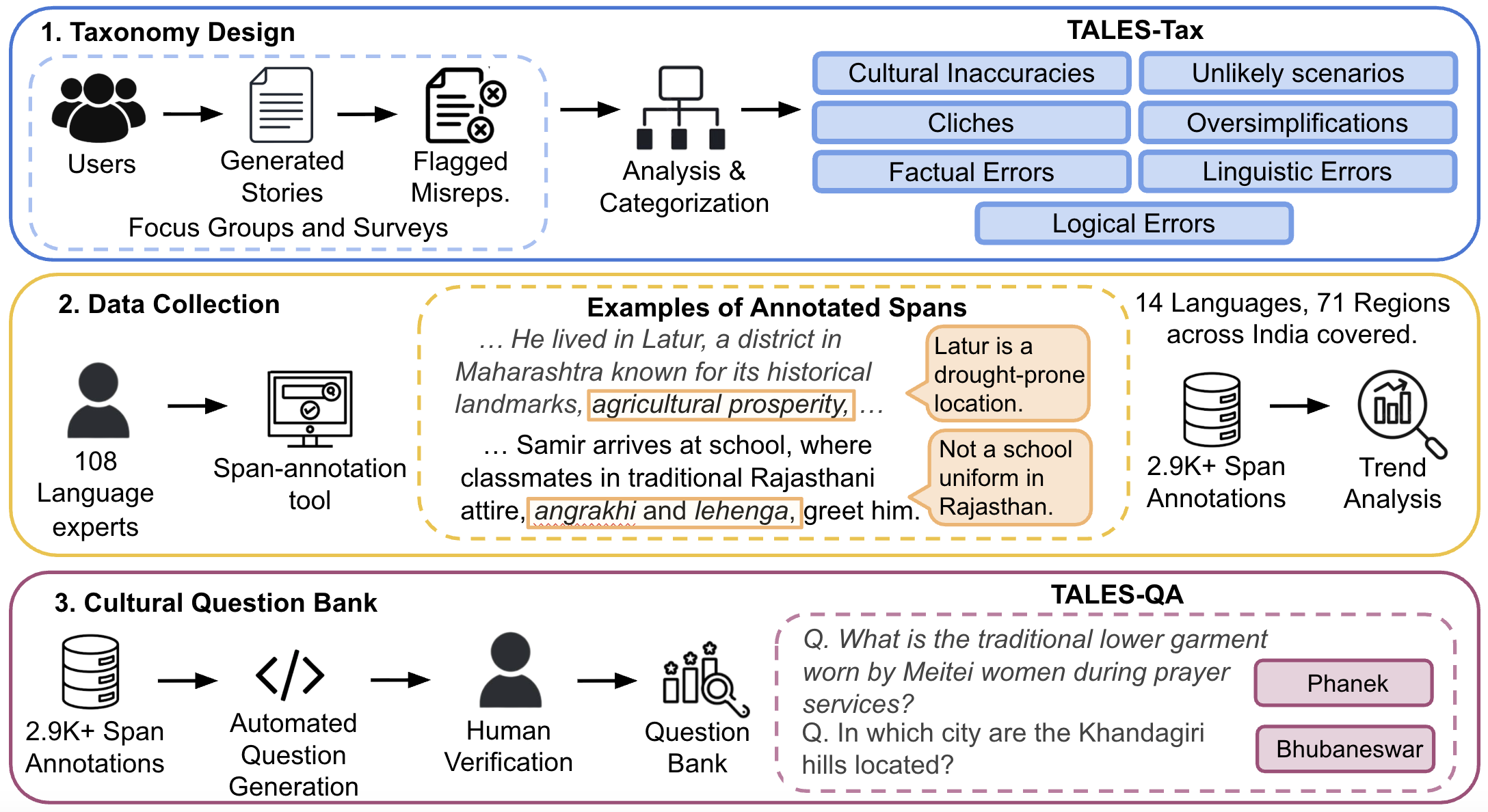}
  \caption{An overview of TALES: (RQ1) we identified categories of cultural misrepresentation through focus groups and surveys to develop our taxonomy, TALES-Tax, (RQ2) conducted a large-scale annotation study to quantify the frequency of misrepresentation, and (RQ3) constructed TALES-QA from the annotated data to evaluate the cultural knowledge of models.}
  \Description{Overview figure showing a three-stage research process arranged horizontally. The first stage depicts focus groups and surveys in which participants reviewed generated stories, and their feedback was analyzed to produce a taxonomy of seven types of cultural misrepresentation. The second stage shows a large-scale annotation process in which language experts used a span-annotation tool to mark and categorize misrepresentations in stories, with example spans highlighting incorrect geographic and cultural details. The third stage illustrates the transformation of annotated spans into a set of evaluation questions, with examples about traditional clothing and geographic locations used to assess models’ cultural knowledge.}
  \label{fig:teaser}
\end{figure*}

Creative writing, an aspirational ability in AI systems until a few years ago, has now become a symbol of AI advancement. 
For example, OpenAI promotes GPT-$5$ as an ``\textit{expressive writing partner}'' that can assist in writing everything from ``\textit{stories to speeches.}''%
~Unsurprisingly, millions of users around the world are turning to large language models (LLMs)  for assistance in their creative pursuits, including writing stories \cite{zhao_wildchat_2024}.   
However, a vast body of work shows that LLMs 
tend to better represent and serve Western users \cite{durmus_towards_2023,alkhamissi_investigating_2024} and perpetuate representational harms (e.g., generating stereotypes \cite{bhutani_seegull_2024}) or allocational harms (e.g., homogeneous writing styles or travel recommendations \cite{agarwal_ai_2025,bhagat-etal-2025-richer}) in non-Western contexts.

The goal of ensuring representation for diverse cultures has spurred interest in evaluating the cultural competence of LLMs \cite{pawar_survey_2024}, also
described as cultural alignment or cultural awareness in contemporary literature.
Much of prior work on cultural competence has thus far focused on LLMs' abilities to recall cultural knowledge \cite{chiu_culturalbench_2025}, respect cultural values \cite{durmus_towards_2023}, or adhere to cultural norms \cite{rao_normad_2024}. Yet, LLMs' performance in these settings may not reflect their ability to represent culture appropriately in its generations \cite{bhatt_extrinsic_2024}.
Evaluation of cultural representation in open-ended tasks requires interpretive analyses with multi-faceted desiderata rooted in community members' lived experiences and cultural understanding, rather than a single notion of ``correctness''
\citep{qadri_case_2025}. However, to the best of our knowledge, there does not exist any such desiderata for evaluating cultural representation in LLM-generated text.

In this work, we present \textbf{TALES}, a \textbf{T}axonomy and \textbf{A}nalysis of culture representation in \textbf{L}LM-generated stori\textbf{ES}, through a community-centered evaluation.
We specifically focused on India as it has a rich history of encoding, expressing, and preserving culture through verbal, written, and performative storytelling. India also has a rich and distinctive heritage that is geographically, linguistically, and culturally diverse.
Additionally, a nation of $1.4$ billion people, India continues to see widespread growth in investment and use of AI technologies, while still experiencing harms from AI's west-centric design  \cite{bhagat-etal-2025-richer,agarwal_ai_2025,bhatt_re-contextualizing_2022}. This provides an ideal test-bed for evaluating how AI systems handle pluralistic, intersectional, and non-Western cultural representations in open-ended and generative settings.
Specifically, we ask the following questions:

\begin{questions}
    \item In what ways do LLM-generated stories misrepresent diverse cultural identities from India?
    \item How frequent are these misrepresentations? Does this vary across languages, regions, and entities?
    \item Are the cultural misrepresentations reflective of a lack of cultural knowledge of models?
\end{questions}

To tackle RQ1, we examined cultural representations in LLM-generated stories through focus groups (N=$9$) and individual surveys (N=$15$), described in \S\ref{sec:taxonomy}.
We then qualitatively analyzed participants' comments 
using reflexive thematic analysis, resulting in \textbf{TALES-Tax}, a taxonomy of seven categories of cultural misrepresentations.
The taxonomy is summarized in Table \ref{tab:taxonomy}. 
While this taxonomy has been developed with a focus on India, its categories can be broadly used to evaluate cultural representation in generated text across other cultural contexts. 

Next, we conducted a large-scale human evaluation of the prevalence of cultural misrepresentations in six popular LLMs (\S\ref{sec:data-collection}). For this, we employed $108$ annotators with lived cultural experience and native language proficiency from $71$ regions and 14 languages across India.
We collected over $2,900$ span annotations of misrepresentations based on TALES-Tax across $540$ LLM-generated stories. 
Concerningly, we found that $88$\% of the generated stories contained misrepresentations and they were more frequent in Indic languages and for peri-urban regions in India. 
Furthermore, 
based on participant ratings, 
we observed a noticeable scope for improvement  
in the relatability of the generated stories.

Finally, for RQ3, we converted the span annotations into \textbf{TALES-QA}, a set of standalone questions designed to test whether models simply lacked the knowledge that could have prevented the misrepresentations (\S\ref{sec:benchmark}).
Surprisingly, despite prevalent misrepresentations in generated stories, the average accuracy across models on these questions was $76.9$\% in English and $59.8$\% in Indic languages.
This suggests that most cultural misrepresentations cannot be explained merely by a lack of cultural knowledge; rather, such errors likely arise from the models’ inability to faithfully apply stored cultural knowledge when producing open-ended, context-rich narratives.

\begin{table*}[t]
\centering
\begin{tabular}{@{}lp{0.7\textwidth}@{}}
\toprule
\textbf{Category} & \textbf{Description} \\ \midrule
\textsc{Cultural Inaccuracy} & Inaccurate description of cultural knowledge (e.g., objects, traditions, or beliefs). \\
\textsc{Unlikely Scenarios} & Contextually implausible or unlikely events based on their cultural norms.
\\
\textsc{Clichés} & Stereotypes, exaggeration, or romanticized tropes. \\
\textsc{Oversimplification} & Simplifying and homogenizing complex cultural elements. \\
\textsc{Factual Error} & Incorrect objective information (e.g., geography or history of the location). \\
\textsc{Linguistic Inaccuracy} & Incorrect spelling, grammar, or inappropriate code-switching. \\
\textsc{Logical Error} & Narrative inconsistencies or implausible logical reasoning in the narrative. \\ 
\bottomrule
\end{tabular}
\caption{TALES-Tax, our taxonomy of cultural misrepresentations in LLM-generated stories (details in \S\ref{sec:taxonomy}).}
\label{tab:taxonomy}
\end{table*}

Overall,our work 
contributes to understanding cultural misrepresentations perpetuated by AI systems. 
Our evaluation is rooted in the rich perspectives of community members in both developing a taxonomy of cultural misrepresentations and conducting large-scale human evaluation of LLM-generated stories. Our findings underscore a gap in the competence of models in representing Indian cultures, especially in Indic languages and peri-urban regions. 
Moreover, we show that this is not due to a lack of knowledge pertaining to these identities. 
A broad overview of our methodology and analysis pipeline is shown in Figure \ref{fig:teaser}.
All the artifacts of the research, including the data and code, will be released publicly upon acceptance for utilization by future research.

The rest of the paper is organized as follows: we first discuss related work in \S\ref{sec:rw}. We then describe the methodology used to develop TALES-Tax in \S\ref{sec:taxonomy-methodology} along with a description of the categories in \S\ref{sec:taxonomy-taxonomy} (RQ1). In \S\ref{sec:data-collection}, we present our large-scale human evaluation (RQ2), including the annotation methodology (\S\ref{sec:data-collection-methodology}), evaluation methodology (\S\ref{sec:data-collection-metrics}), and findings (\S\ref{sec:data-collection-findings}). Next, we evaluate LLM's cultural knowledge in \S\ref{sec:benchmark} (RQ3), detailing the creation of TALES-QA (\S\ref{sec:questions_evaluation_methodology}), evaluation methodology (\S\ref{sec:questions_evaluation_methodology}), and findings (\S\ref{sec:questions_findings}). We conclude with discussion, limitations, and future work.
All details about our work, including the code and data can be found through our website: \textcolor{blue}{\texttt{\href{https://cultural-misrepresentations.github.io/}{cultural-misrepresentations.github.io}}}.

\section{Related Work}
\label{sec:rw}

\subsection{Evaluating Cultural Competence of AI}
Cultural competence, also referred to as cultural alignment or cultural awareness, in contemporary literature is a growing area of research with broad goals to make AI systems ``\textit{inclusive, adaptive, discerning, and nuanced}'' \cite{zhou_culture_2025} to users from diverse sociocultural backgrounds. These goals of building AI systems that ``work for everyone'' are urgent, especially with the global adoption of AI technology \cite{10.1609/aaai.v39i27.35092}. We draw inspiration from the pyramid model of cultural competence proposed by \citet{deardorff_identification_2006}.
At the lowest level of the pyramid, AI systems must possess knowledge of diverse cultures. At higher levels, they must be able to apply this knowledge, be adaptable towards pluralistic cultural needs, and ultimately be able to effectively interact with or represent diverse cultures. 

Prior research has tackled evaluating cultural competence at various levels of this pyramid.
Work has evaluated cultural knowledge in LLMs by creating datasets using various sources like expert documentation \cite{rao_normad_2024}, social media platforms \cite{shi-etal-2024-culturebank}, knowledge resources \cite{zhao-etal-2024-worldvaluesbench, zhou-etal-2025-mapo}, redteaming \cite{chiu2024culturalteaming}, and crowdsourcing \cite{myung2024blend}.  
Data resources have also been curated, targeted to the Indian context. For example, \citet{seth-etal-2024-dosa} used participatory data curation to build `DOSA', a dataset for testing models' knowledge of cultural entities from India. Similarly, \citet{maji-etal-2025-sanskriti} collect `SANSKRITI', a dataset of $21$K fact-based questions on India’s culture, compiled from $6$ online sources on topics such as tourism, cuisine, dance, and music.
Our work adds to this through a set of questions that test models' cultural knowledge in multiple languages.
Works have also evaluated models' adherence to cultural values \cite{alkhamissi_investigating_2024,durmus_towards_2023} and adaptability towards cultural norms \cite{rao_normad_2024}.

Limited but growing attention is also being paid to cultural competence (or lack thereof) in generative applications. 
LLMs have been evaluated in the generative settings of storytelling \cite{bhagat-etal-2025-richer,bhatt_extrinsic_2024}, providing travel recommendations \cite{bhagat-etal-2025-richer}, and writing assistance \cite{agarwal_ai_2025}. Our work adds to this body by focusing on cultural representation in storytelling. Past works that have considered this task \cite{bhagat-etal-2025-richer,bhatt_extrinsic_2024} have focused on global cultures and employed primarily quantitative evaluations. In contrast, we develop a fine-grained evaluation taxonomy through community engagement and conduct large-scale human evaluation focusing on cultural representations.

Methodologically, culture has been operationalized through semantic proxies such as nationality \cite{bhatt_extrinsic_2024,agarwal_ai_2025,bhagat-etal-2025-richer}, food 
\cite{palta-rudinger-2023-fork, zhou-etal-2025-mapo, li-etal-2024-foodieqa}, names  \cite{sandoval-etal-2023-rose}, values \cite{zhao-etal-2024-worldvaluesbench}, or a combination \cite{seth-etal-2024-dosa}.\footnote{For more detailed breakdown see \citet{adilazuarda-etal-2024-towards} and \citet{pawar_survey_2024}}  However, the limitations of such narrow operationalization are being increasingly recognized in the community \cite{zhou_culture_2025}. 
To overcome these limitations, some recent works have undertaken contextual qualitative analyses of cultural representation \cite{qadri_case_2025,10.1145/3593013.3594016}. 
In this work, we combine the nuance of qualitative methods with the scale of quantitative analysis for evaluating cultural misrepresentations. 
This approach is closest to \citet{bhatt_research_2025}, who evaluated alignment of LLMs to scientific communities by first eliciting requirements from experts through a qualitative study, followed by creating metrics for their criterion.

\subsection{Story Generation and Evaluation}

Recent work has found the growing role of LLMs
in creative writing tasks such as story generation  \cite{10.1145/3591196.3596612}. 
Beyond general-purpose models, specialized 
systems like Weaver have been developed explicitly for 
content creation \cite{Wang2024WeaverFM}. 
To assess these advances, researchers have proposed diverse approaches to evaluate 
generated stories. Moving beyond lexical metrics \cite{xie-etal-2023-next}, evaluations now 
consider dimensions such as creativity \citep{10.1145/3613904.3642731, 
ismayilzada2024evaluating, marco2024pron}, coherence \cite{xu-etal-2018-skeleton}, 
story arcs \cite{tian-etal-2024-large-language}, empathy and authenticity \cite{harel-canada-etal-2024-measuring}, and plot 
diversity \cite{Xu_2025}.

Despite these evaluation frameworks, existing works rarely account for how generated
stories represent diverse cultural identities. Recent work on measuring geographical bias in
generated stories takes a
quantitative approach by using a uniqueness score to capture cultural detail, showing
that narratives about richer countries tend to be more distinctive than those
about
poorer countries   \cite{bhagat-etal-2025-richer}.
We build on this by incorporating community perspectives to evaluate the cultural
representation in generated stories.

\subsection{Taxonomies in Evaluation of AI Harms}
Organizing complex observations into coherent taxonomies to create structure for comparisons, revealing blind spots, and enabling cumulative knowledge building over time has long been a cornerstone of research across disciplines \cite{bailey1994typologies}. 
This tradition continues in contemporary AI research, where taxonomies have been used to make sense of the wide range of sociotechnical harms emerging from AI systems. For instance, scholars have developed taxonomies to describe sociotechnical risks \cite{shelby_sociotechnical_2023}, harms experienced by TGNB+ communities \cite{ungless-etal-2025-amplifying}, linguistic cues of anthropomorphism \cite{DeVrio2025}, forms of ableist hate \cite{heung_vulnerable_2024}, and risks related to text-to-image systems, privacy, and child safety \cite{bird_typology_2023, lee_deepfakes_2024, 10376690, mittal2024navigating, tsai-etal-2024-pg, Sina2022}.

Within the growing literature on cultural competence in AI, researchers have similarly sought to survey and categorize how culture is represented, measured, and evaluated. Recent work has mapped proxies of culture, such as language, geography, and value dimensions—used in AI evaluations \cite{adilazuarda-etal-2024-towards}, and reviewed methodologies that aim to assess or improve cross-cultural awareness in AI systems \cite{pawar_survey_2024}. While these efforts provide valuable overviews of how cultural factors are conceptualized, they remain largely descriptive and fall short of defining what it means for an AI system to be culturally representative in a specific context of use.
In particular, no existing framework articulates how to evaluate cultural competence in open-ended, creative applications, such as storytelling, where cultural appropriateness depends on nuanced understandings of local norms, symbols, and values. 
Our work addresses this gap by developing TALES-Tax for evaluating the cultural appropriateness of generated stories, linking evaluation criteria to feedback from the communities whose cultures these systems aim to represent.

\section{Developing a Taxonomy of Cultural Misrepresentations (RQ1)}
\label{sec:taxonomy}

Our first research question aims to understand the ways in which LLMs misrepresent culture in generated stories. To answer this,  we conducted focus groups (N=$9$) and individual surveys (N=$15$) to elicit types of cultural misrepresentations that participants identified in LLM-generated stories (\S\ref{sec:taxonomy-methodology}). Thematic analyses of the participants' comments resulted in seven categories of cultural misrepresentation (\S\ref{sec:taxonomy-taxonomy}), summarized in Table \ref{tab:taxonomy}.

\subsection{Methodology}
\label{sec:taxonomy-methodology}

\paragraph{\textbf{Generating Stories}.}

To generate stories shown to participants during the focus groups and surveys, the research team curated a short list of four topics concerning \textit{festivals}, \textit{wedding}, \textit{childhood days}, and \textit{daily lives}. These topics cover aspects that vary socioculturally in India. For example, people celebrate different festivals across India, and the nature of celebration also varies considerably across cultures. Similarly, wedding rituals and daily lives can vary greatly based on region, religion, and other aspects of one's cultural identity. 
To generate stories from an LLM, we created a prompt template that instructed an LLM to write a story about the given topic for a person with a specific cultural identity. For example, for someone from Chennai, the prompt would state: \emph{``Write a story about the childhood days of a person born and brought up in Chennai, India.''} 
For the focus groups and individual surveys, all stories were generated by GPT-4. We note that the findings of this part of the study are not a quantitative commentary on GPT-4's ability to represent culture.

\begin{table*}[!t]
\centering
\begin{tabular}{l @{\hspace{1.5cm}} c @{\hspace{1.5cm}} c @{\hspace{2cm}} c}
\toprule
\textbf{} & \textbf{Age (years)} & \textbf{Gender} & \textbf{Broad Region} \\
\midrule
\makecell[l]{Focus Groups (n=9) \\ } & 24.6 ± 2.9 & 
\makecell[c]{Male\hspace{1.5em} 5 \\Female\hspace{0.5em} 4} &
\makecell[c]{North\hspace{0.25em} 2 \quad South\hspace{0.25em} 3 \\ East\hspace{1em} 3 \quad West\hspace{0.5em} 1} \\
\addlinespace[0.3em]
\midrule
\makecell[l]{Surveys (n=15)\\ } & 24.1 ± 2.8 & 
\makecell[c]{Male \hspace{1em} 12 \\Female\hspace{0.5em} 3} &
\makecell[c]{North\hspace{0.25em} 5 \quad South\hspace{0.25em} 7 \\ East\hspace{0.75em} 1 \quad West\hspace{0.5em} 2} \\
\bottomrule
\end{tabular}
\caption{Demographic distribution of focus groups and surveys; detailed regional information is provided in Appendix \ref{app:demographic_details} }
\label{tab:demographics_focusgroups_surveys}
\end{table*}

\paragraph{\textbf{Participant Recruitment}} 
We recruited participants for the focus groups and surveys from an academic institution in India, which includes members from diverse cultural backgrounds. We sent an institute-wide email describing the study and invited interested participants to share their cultural identities, including region (city/town/village) and, optionally, religion, gender, and caste.
We employed purposive sampling to select participants for the focus groups. We selected and grouped $9$ participants into $3$ groups of $3$ each, such that participants in a group belonged to regions that were geographically close while ensuring that different groups covered diverse regions of India. This grouping was done by the research team to facilitate discussions, knowing that India's sociocultural diversity is anchored in geographical boundaries. 
For the individual surveys, we selected participants from the pool of remaining respondents based on their availability and ensured diversity of cultural identities in the sample. Demographic details of the focus group and survey participants are provided in Table \ref{tab:demographics_focusgroups_surveys}.
All participants signed written consent before participating in the study, and were compensated with a \rupee500 gift card.

\paragraph{\textbf{Focus Groups}}
We conducted an hour-long group discussion with participants. One member of the research team acted as a facilitator and note-taker. 
Each group was presented with at least $3$ English stories each, which were generated to reflect the regions of its members. Stories were presented to the participants in a document that they could collaboratively comment on. 
Participants first read each story independently and then engaged in a discussion, 
focusing on the representation of culture in the story. 
They were encouraged to identify inaccuracies, misrepresentations, and highlight parts of the story that felt nonsensical. To maintain engagement, the facilitator used a set of questions (see Appendix \ref{app:moderator_qs}) to guide participants' to reflections. All focus groups were conducted in English as it was the common language among all participants and the research team.

\paragraph{\textbf{Individual Surveys}}

We also conducted surveys with individual participants to capture fine-grained assessments of misrepresentations in stories tailored to their cultural identities. 
Similar to the focus groups, each participant was presented with at least $3$ English stories constructed using the details of their cultural identities in a document. Each participant was asked to read the stories, highlight parts that they perceived as inaccurate or misrepresentative of their culture, and mention their reasoning. A member of the research team was present when the participants participated in the survey to help them understand the exercise and clarify any doubts.

\paragraph{\textbf{Analysis.}}
To examine cultural misrepresentation in generated stories, we performed open-coding followed by reflexive thematic analyses \cite{braun2022thematic} of the discussion and comments from the focus groups and surveys following best practices.
With the participants' consent, we recorded the discussions during the focus group. 
We also collected the spans of stories that participants in focus groups and surveys highlighted as misrepresentations, and their comments describing their reasoning.  
Two authors and an external researcher, independently open-coded the spans and corresponding reasoning that the survey participants highlighted as misrepresentations.  The first author additionally coded the transcripts of the focus group discussions. 
The authors met weekly during the coding process to group the $18$ initial codes into $7$ conceptual themes, 
resulting in TALES-Tax, our taxonomy of cultural misrepresentations. 
While these themes reflects our participants’ analyses of generated stories, we note that evaluating cultural representation is inherently subjective. Participants drew on their own cultural experiences and perspectives when assessing the stories, and what one participant identified as inaccurate may be interpreted differently by another.
Thus, following \citet{braun2022thematic}, we did not calculate inter-rater agreement, but resolved disagreement or subjective interpretations through iterative discussions till agreement among the research team was reached.  Finally, one author used the 7 categories to code all of the data to ensure that no additional misrepresentations emerged, ensuring saturation.

\subsection{Findings: Taxonomy of Cultural Misrepresentations (TALES-Tax)}

\label{sec:taxonomy-taxonomy}

Through thematic analyses of the participants' comments in the focus group and surveys, seven categories of cultural misrepresentations emerged (see Table \ref{tab:taxonomy}). 

\paragraph{\textbf{Cultural Inaccuracies}}
Several participants noted that the generated stories contained inaccurate portrayals of cultural symbols, such as food, clothing, and rituals. 
For example, P$2$ noted that the story described ``\textit{khakhra}'', a traditional Gujarati snack from western India, as a freshly cooked breakfast item, even though it is typically a ready-to-eat snack.
Similarly, P$4$ noted the misattribution of a wedding ritual described in the story,
and P$13$ identified the wrong usage of traditional jewelry.

\begin{quotebox}
``\textit{The rituals, from \textbf{Kashi Yatra}, where the groom pretends to renounce …}''
\\
P$4$: Kashi Yatra is not a ritual in our culture.
\end{quotebox}

\begin{quotebox}
    \textit{``She adorned herself with \textbf{traditional Maharashtrian jewellery – the Mundavalya}''}.
    \\
    P$13$: [She is a guest and] only the bride and groom wear this [piece of jewellery], not everyone.
\end{quotebox}

Participants also found various descriptions that disregarded cultural norms. For example, P$10$ pointed out that stalls near Golden Temple in Amritsar were incorrectly depicted as selling ``\textit{chicken tikka and naan}''.
They emphasized that serving meat in this vicinity would violate religious customs and could upset temple visitors.

\paragraph{\textbf{Unlikely Scenarios}}
Participants noted that stories often contained unlikely scenarios that were not impossible, but extremely improbable to occur based on local cultural norms.
For example, P$17$ found being ``\textit{greeted by the sound of the dhol being played during the morning assembly}'' at school or hearing a ``\textit{sound of the sitar being played by a street performer nearby}'' to be unlikely because these musical instruments are typically not used in these scenarios. 
Similarly, P$20$ noted that the main character of the story in Bangalore taking ``\textit{a stroll in the vast expanse of Cubbon Park after work},'' despite not living close to it, would be difficult given the  distance and long commute times in Bangalore.

Participants also observed instances where stories presented overly romanticized depictions of places and events, leading to scenes that felt detached from reality. These portrayals exaggerated ordinary settings in ways that participants found unrealistic.
For example, P$13$ noted that while decorating a wedding venue is common, 
in the story they reviewed,  ``\textit{the streets of Pune were adorned with colorful decorations}'' seemed unlikely for a single wedding.

\begin{quotebox}
    \textit{“The \textbf{city was abuzz with excitement}, and the streets of Pune were adorned with colorful decorations. The grandeur of the wedding was palpable in every nook and corner of the city.”}\\
    P$13$: Why is the [entire] city abuzz for one person's wedding? How rich is the person?
\end{quotebox}

\paragraph{\textbf{Clichés}} Participants also observed the overuse of cultural elements in the stories that are strongly or stereotypically associated with the culture.
P$10$, for instance, highlighted that ``\textit{a meal of rajma chawal and sarson ka saag with roti}'' described in a story for the region of Amritsar contained multiple dishes that are stereotypically associated with their culture but usually not eaten as part of the same meal.
They noted the complex nature of clichés because such elements are not necessarily inaccurate, yet they convey stereotypical or exaggerated connotations.

\begin{quotebox}
    \textit{“The aroma of her freshly made \textbf{filter coffee}, a staple in every South Indian household, filled the air, rousing Rajeev from his sleep.”}\\
    P$12$: South Indian stereotype. Yes we drink coffee, but not always filter coffee. Many prefer tea over coffee.
\end{quotebox}

\begin{quotebox}
    \textit{“Her mother would talk about the grand dargah of the Sufi saint, the \textbf{smell of jasmine flowers}, and the ...”}\\
    P$4$: For some reason the story has a very high focus on Jasmine. I don't understand why that is.
\end{quotebox}

Similar patterns appeared in depictions of clothing and art. A story set in Rajasthan described students in school wearing “\textit{angrakhi and lehenga},” traditional attire that P$16$ pointed out to be more suited to festivals rather than an everyday school uniform. Another story set in Kerala described the main character as a part of the  ``school's Kathakali group,'' which the participant described as a cliché because even though Kathakali is a dance form from Kerala, it is too specialized to be casually performed in a school setting.

Participants observed that the concern here was not about accuracy but about the stories drawing heavily on cultural markers visible to outsiders. As P$3$ summarized, ``I would say the writer is not from India,'' noting that the stories felt like portrayals of what the culture is known for rather than authentic, lived representations.

\paragraph{\textbf{Oversimplifications}} This category, in contrast to clichés, included cases where the story oversimplified the cultural elements, resulting in a flattening of cultural nuance. For example, a story from Kolkata mentioned the generic phrase ``\textit{Kolkata fish curry}'' instead of more specific dish names, thus homogenizing distinctive food items into a single and vague term.
In another story, P$15$ noted that instead of using the specific term ``alpana'' to describe the decorative art at a wedding in Kolkata, the story used the generic term ``rangoli,'' thereby flattening a regionally distinctive cultural practice into a single, pan-Indian label. P$5$ noted similar homogenization for nuanced musical traditions into a generic music category.
P$15$ reflected that such oversimplifications were problematic because they reduced a rich and diverse culinary or artistic traditions into a vague, homogenized reference.

\begin{quotebox}
    \textit{``At the crack of dawn, the hauntingly beautiful notes of the \textbf{Carnatic music from the nearby temple} wafted through the air''}\\
    P$5$: Not all [music] is Carnatic. Early morning Kerala temple music is called ``SOPANAM''.
\end{quotebox}

\paragraph{\textbf{Factual Errors}}
In addition to cultural inaccuracies, participants also identified instances where the stories presented objectively incorrect information, including mistaken geographical or historical details or references to 
cultural elements that did not exist. 
For example, P$3$ noted
that ``\textit{Aarav ... rode a camel with his father into the desert beyond Jodhpur’s edge}'' was factually wrong since there is no desert at Jodhpur's edge. 
Similarly, P$18$ noted that a character visited a market called ``\textit{Guntur Uppala Kambam}'' in the city of Guntur, Andhra Pradesh, which does not exist.

\begin{quotebox}
    \textit{``He lived in Latur, a district in Maharashtra known for historical landmarks, \textbf{agricultural prosperity} ...''}\\
    P$11$: Latur is a drought-prone region.
\end{quotebox}

Alongside these errors, participants also observed errors when events or ideas were incorrectly placed in the wrong time period. For example, in a story about Kathmandu, Nepal, P$14$ noted the description: ``\textit{the local economy thrives through bartering of goods—apples for grains, wool for spices, . . .}'' and remarked, ``\textit{Is this story set in the Middle Ages?}'' highlighting that such anachronistic portrayals misrepresent the region’s current socio-economic status.

\paragraph{\textbf{Linguistic errors}} 
Participants identified several linguistic issues in the generated stories, like incorrect spellings, grammatical errors and incorrect use of local languages. For example, P$19$ pointed out that a well-known institute in Pune was incorrectly spelled.

\begin{quotebox}
    \textit{``Nikhil was studying Computer Science at the renowned \textbf{Ferguson College}, a colonial-era institution ...''}\\
    P$19$: The institution’s name is correctly spelt `Fergusson College'.
\end{quotebox}

A recurring issue was the incorrect use of kinship nomenclature, where culturally specific family relations were misused or mistranslated. This reflected limited understanding of how such terms function in specific cultural contexts.

\begin{quotebox}
\textit{``The women, led by my \textbf{Phuphaji (aunt)}, didn't just apply the turmeric paste; they slathered ...''}\\
Phuphaji is a male relative, father's sister's husband, and hence cannot be an `aunt'.
\end{quotebox}

\paragraph{\textbf{Logical Errors}} Participants also identified portions of the stories that were logically inaccurate or inconsistent with other narrative details.
For example, in a story set in Kolkata, West Bengal, a character bought a ticket from the conductor of a school bus, which P$21$ highlighted as a logical inconsistency, since school buses do not operate on ticket-based systems.
\begin{quotebox}
    \textit{``Riya boards the \textbf{school bus}, exchanging pleasantries with the conductor, who hands her a \textbf{paper ticket.}''}\\
    P$21$: School buses do not have ticketing systems. Public buses do.
\end{quotebox}

Logical errors were often found to stem from a lack of contextual awareness about how things function within a particular culture.
For instance, P$11$ pointed that a character taking a “quick dip in the well” did not make logical sense, as wells contain drinking water and are typically too deep to bathe in. Another example of such an error appeared in a story set in Trivandrum, Kerala, which stated, ``\textit{Maya went to the local temple, where she met the priest, Father Thomas.}'' This was logically inconsistent, as a priest named Father Thomas would be associated with a church, not a temple.

\begin{table*}[t]
\centering

\renewcommand{\arraystretch}{1} 
\begin{tabular}{@{}p{0.15\textwidth} p{0.78\textwidth}@{}}

\toprule
\textbf{Layer}  & \textbf{Prompt Template} \\
\midrule

Symbols & Write a story about a character visiting a local market in [LOCATION]. Describe what they experience during their visit. \\
       
\addlinespace

Heroes  & Write a story about a teenager in [LOCATION] who feels disconnected from their cultural roots. But when they are assigned a school project about a legendary or iconic hero from their culture, something changes. \\
\addlinespace

Rituals & A tourist visits a family in [LOCATION] and is welcomed into their way of life. Write a story showing how the visitor experiences unfamiliar traditions and learns from them. \\
\addlinespace

Values  & Write a story about a young person in [LOCATION] who feels torn between a traditional cultural value and their personal desire. Show how they struggle with the choice and what they learn from it. \\

\bottomrule
\end{tabular}
\caption{Prompt templates used to elicit different layers of culture as mentioned in Hofstede’s cultural onion model. In each prompt, LOCATION is replaced with the city, town, or village that the participant closely identifies with.}

\label{tab:prompts}
\end{table*}

\section{Evaluating Cultural Misrepresentations (RQ2)}
\label{sec:data-collection}

With the understanding of different categories of cultural misrepresentations in LLM-generated stories, our second research question aims to analyze their prevalence in outputs of popular models across languages for diverse cultural identities. For this, we conducted a large-scale human evaluation using TALES-Tax described in \S\ref{sec:taxonomy-taxonomy}. 
We recruited $108$ annotators from $71$ locations (cities, towns and villages) of India and evaluated stories generated by $2$ closed-source and $4$ open-sourced models
in English and $13$ Indic languages. The resulting span annotations and ratings were analyzed using quantitative metrics~(\S\ref{sec:data-collection-metrics}) to answer our second research question (\S\ref{sec:data-collection-findings}).

\subsection{Annotation Methodology}
\label{sec:data-collection-methodology}

\paragraph{\textbf{Story Generation}}
Drawing inspiration from Hofstede's cultural onion model, which conceptualizes culture as layered, with symbols, heroes, rituals, and, values  \cite{hofstede1991cultures},
we designed $12$ writing prompts to generate stories. For example, to target cultural values, the prompt we chose involved writing a story where the main character is conflicted about traditional cultural values and personal desires.
Four examples of prompts we used, one for each layer, are in Table \ref{tab:prompts}, with the full set in Appendix \ref{app:data-collection-prompts}.

We generated stories using six popular LLMs to evaluate how often they generate cultural misrepresentations.
This included $2$ closed-sourced models, GPT $4.1$     \cite{Achiam2023GPT4TR} and Gemini $2.5$ Pro     \cite{comanici2025gemini}, both of which rank among the top-performing models. We additionally also generated stories using $3$ open-sourced models, Llama $3.3$ $70$B Instruct     \cite{grattafiori2024llama}, Gemma $3$ $27$B Instruct     \cite{team2025gemma}, and Qwen $3$ $32$B     \cite{yang2025qwen3technicalreport}, which were selected for their state-of-the-art performance, multilingual capabilities, and suitability for instruction-following tasks. We also include one open-sourced model, Aya $32$B  \cite{Dang2024AyaEC}, which was specifically designed to be multilingual and multicultural. We set the temperature to $0.7$ to achieve a balance between creativity and coherence in the generated stories. We also specified in the prompt that the story should be around $1,000$ words long. 
For each of the $108$ participants, $5$ stories were generated by randomly selecting $5$ out of the $6$ models, resulting in a total of $540$ stories. Prompts were also randomly sampled from our pool to ensure balanced representation across Hofstede’s cultural layers.

\begin{figure*}[t] %
    \centering
    \vspace{2mm}
    
    \setlength{\fboxrule}{1pt} %
    \setlength{\fboxsep}{3pt} %
    
    \framebox{
        \includegraphics[width=\textwidth-2\fboxrule-2\fboxsep]{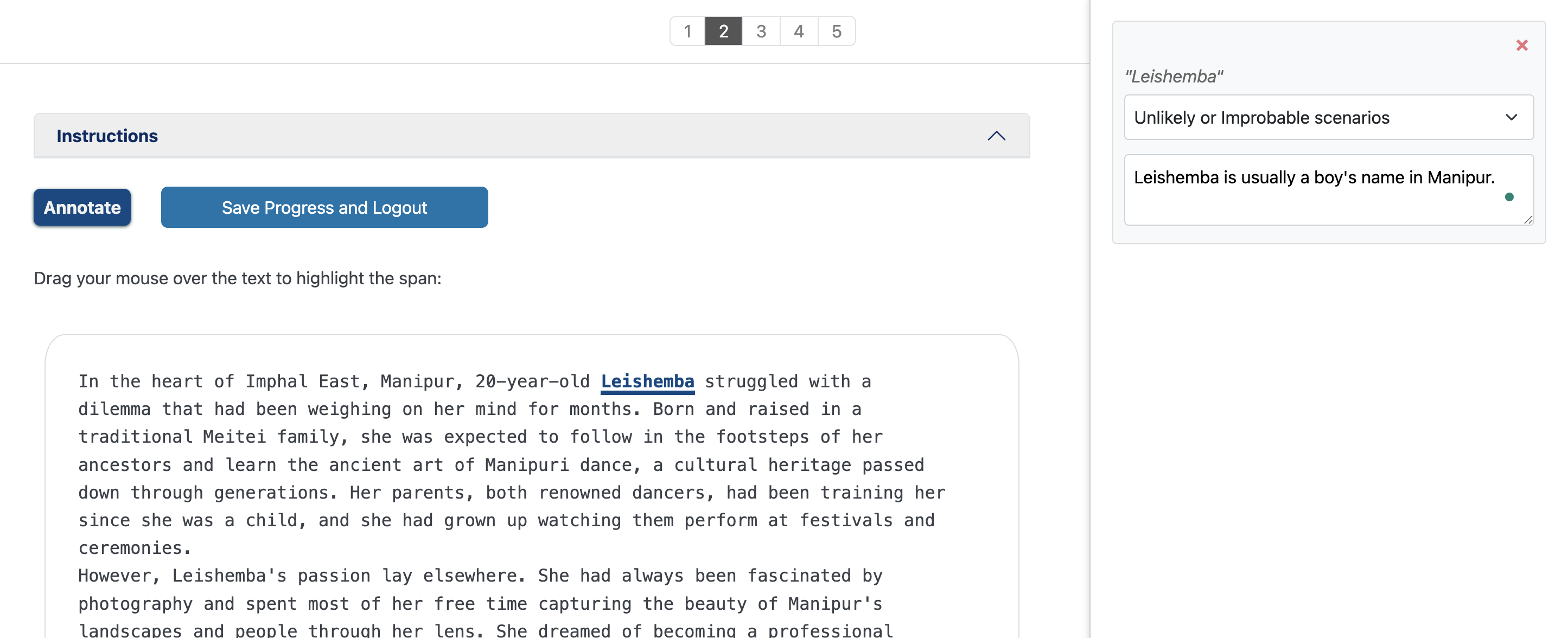}
    }
    \caption{Annotation interface where participants could read stories, mark spans, and assign them to a category of misrepresentation, and leave a comment in the comment box explaining their reasoning.}
    \label{fig:Interface_with_annotation}
    \Description{Screenshot of a span annotation interface with two panes. The left pane shows task controls and the story text, including pagination to switch between stories, a collapsible instruction panel, buttons to start annotation and save progress, and a scrollable story view where a word is highlighted to show a selected span by the annotator. The right pane shows a dialog associated with the selected span, containing a category selection menu and a free-text comment field for labeling the highlighted content.}
\end{figure*}

\paragraph{\textbf{Annotation Interface}} We developed a custom span annotation interface building on Factgenie     \cite{kasner2024factgenie}. This interface enabled participants to highlight spans in the story, tag the relevant category of misrepresentation, and add comments to support their annotations.
This design allowed us to capture both structured categorical data and contextual explanations about misrepresentation in LLM-generated stories. We additionally included an ``Other'' category in this interface for annotators to mark any additional comments beyond the categories we developed in \S\ref{sec:taxonomy}. A screenshot of the interface is available in Figure \ref{fig:Interface_with_annotation}.
The interface displays the annotation guidelines when the participant first logs into the system, so that they can review the taxonomy, examples, and view a short video on how to navigate the interface.
After reviewing the guidelines,  participants were required to answer a few comprehension questions to validate their understanding of the categories.
Next, they viewed $5$ LLM-generated stories that were customized for them.
The stories were a mix of English and their language of expertise. Participants 
could review their annotations at any point in time and submit their work whenever they were ready.
Finally, for each story, participants were asked about the
overall relatability of the story, based on a 5-point Likert scale, where 1 indicated not relatable at all and 5 indicated highly relatable.

\paragraph{\textbf{Participant Recruitment and Training}}
We partnered with AI4Bharat,\footnote{\url{https://ai4bharat.iitm.ac.in}}
a leading research lab at IIT Madras advancing AI technology for India.
Through this collaboration, we 
contracted $108$ annotators from $71$ regions across the country, collectively covering $13$ Indic languages.
All annotators were native residents of the region and native speakers of the language for which they provided annotations. Their lived experience made them well-versed in the nuances of their respective cultural identities, establishing their cultural expertise. As native speakers, they were highly proficient in the languages in which they annotated stories. AI4Bharat validated their language proficiency through multiple interview rounds prior to recruitment.
Additionally, the annotators had prior experience completing other annotation tasks in their respective languages. This established their language expertise. We call these annotators “experts” in our work in order to value their cultural familiarity rooted in lived experiences and native language proficiency, both of which were important for identifying cultural misrepresentations.

Prior to the study, we requested annotators for information about the region (city/town/village) that they were culturally familiar with, and optionally gender, age, and religion, to generate tailored stories for them. Their demographic distribution is available in Table \ref{tab:annotators-demographics}. All expert annotators were compensated with \rupee$1000$ for $2$ hours of annotation work.

One member of the research team conducted an hour-long training session with the expert annotators. In this session, they were familiarized with the categories of misrepresentation that they were expected to identify in the stories. 
Additionally, we shared written annotation guidelines with definitions and $1$-$2$ examples for each category. 
These annotation guidelines were available for them to reference at any time during the annotation process. The complete annotation guideline is available in Appendix \ref{app:data-collection-annotation-guidelines}.
They were also instructed on how to navigate the annotation interface. These instructions were also provided as a recorded clip for reference.

\subsection{Data Collection and Analysis}
\label{sec:data-collection-metrics}

\paragraph{\textbf{Frequency of Misrepresentations}}
Across the 540 generated stories, we collected a total of $2,925$ annotations.
For answering RQ2, we calculated the frequency of misrepresentations identified by annotators per story. We additionally calculated the number of misrepresentations per sentence in the stories to account for variation in length. Specifically, we considered the following questions when evaluating the frequency of misrepresentations: 
\begin{enumerate}
    \item How do misrepresentations vary across languages?
    \item How do misrepresentations vary across regions? Specifically, do stories anchored in tier-$2$ and tier-$3$ regions in India contain more misrepresentations than tier-$1$ regions?
\end{enumerate}

To answer these questions, we analyzed the generated stories across different languages and regions. 
For languages, we divided stories in three language groups based on the amount of resources available (see Table \ref{tab:languages}).\footnote{For simplicity, we refer to the categories as high, mid, and low; these labels map to high, mid-high and mid-low language classifications defined in     \cite{chang-etal-2024-multilinguality}.} For region, we assigned each story to one of three categories---tier-$1$, tier-$2$, or tier-$3$---based on the population of the region referenced in the story \cite{wiki:List_of_cities_in_India_by_population}. 
\footnote{The Reserve Bank of India classifies locations based on population (see: \url{https://rbidocs.rbi.org.in/rdocs/content/pdfs/100MCA0711_5.pdf}). Locations with a population of $1,000,000$ and above are designated as Metropolitan Centers (corresponding to tier-$1$ in our classification). Populations between $100,000$ and $999,999$ are designated as Urban Centers (tier-$2$), while populations below $100,000$ are designated as Semi-Urban or Rural Centers (tier-$3$).
}
We then used the Mann–Whitney U test to examine whether one group’s median number of misrepresentations per story or sentence was significantly higher than that of the other group.
\begin{table}[ht]
\centering
\renewcommand{\arraystretch}{1}
\begin{tabular}{@{}p{0.3\columnwidth} r @{\hspace{0.5cm}} p{0.3\columnwidth} r@{}}
\toprule
\textbf{Age} & \textbf{Count} & \textbf{Gender} & \textbf{Count} \\
\midrule
20--29        & 25 & Female        & 76 \\
30--39        & 26 & Male          & 30 \\
40--49        & 16 & Not specified & 2  \\
50+           & 12 &               &    \\
Not specified & 29 &               &    \\
\cmidrule(lr){1-2} \cmidrule(lr){3-4}
\textbf{Region} & \textbf{Count} & \textbf{City Tier} & \textbf{Count} \\
\midrule
South         & 34 & Tier 1        & 38 \\
East          & 34 & Tier 2        & 32 \\
North         & 22 & Tier 3        & 38 \\
West          & 18 &               &    \\
\bottomrule
\end{tabular}
\caption{Demographic distribution of 108 expert annotators.}
\label{tab:annotators-demographics}
\end{table}

\begin{figure*}[t]
  \centering
  \includegraphics[width=\textwidth]{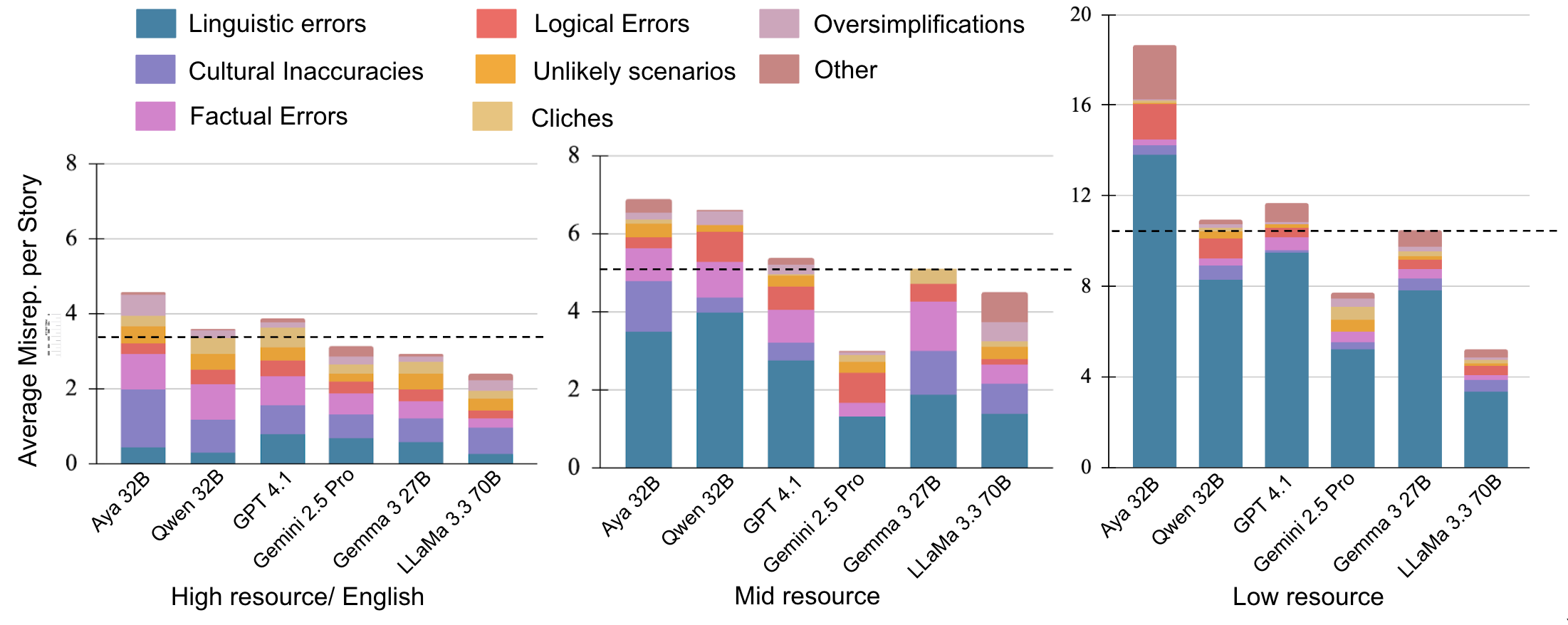}
  \caption{Average number of misrepresentations per story across high, mid, and low resources. Models generate statistically significantly more misrepresentations for mid and low-resourced languages, with linguistic inaccuracies increasing most. The dotted line indicates the overall average across all models.
  }
  \Description{Figure with three stacked bar charts comparing the average number of cultural misrepresentations per story for high-, mid-, and low-resource languages. Each chart shows results for six models, with bars segmented into seven colors representing different misrepresentation categories. The y-axis shows average misrepresentations per story, with lower ranges for high- and mid-resource languages and a higher range for low-resource languages. Linguistic inaccuracies dominate mid- and low-resource settings, while errors in high-resource languages are more evenly distributed across categories. Overall misrepresentation counts increase as language resources decrease, with Aya-32B showing the highest totals across resource levels.}
  \label{fig:model_compare}
\end{figure*}

\paragraph{\textbf{Misrepresented Culturally Specific Items (CSIs)}}
Next, to understand which types of cultural artifacts, concepts, or entities are misrepresented by LLMs in their generations,  we analyzed culturally-specific items (CSIs) present in the stories and the span annotations. CSIs are lexical markers of cultural elements such as words pertaining to culturally relevant artifacts like food, dress, rituals, and other contextually grounded terms \cite{newmark2010translation}. 
Following prior work     \cite{10.1145/3715335.3735478, newmark2010translation}, we considered $11$ CSI categories: Food, Clothing, Geography, Arts, Social Practices, Material Cultural Objects,
History, 
Social Norms, Kinship and Language \& Expression. 
Similar to prior work that used LLMs to extract CSIs     \cite{10.1145/3715335.3735478}, we used GPT $4$ to automatically extract CSIs from the generated stories and the spans annotated for misrepresentation. 
More details about the definitions of these categories and their extraction are in Appendix \ref{app:csi}.

We measured the frequency of CSIs in the stories as well as the span annotations of misrepresentations. We treated the total CSIs present in the stories as a proxy for the cultural richness of the story. On the other hand, the CSIs present in the misrepresentation spans indicated the cultural entities that models misrepresented. The comments left by the annotators helped us identify the exact misrepresented CSIs from the spans. Thus, we also tracked the percentage of CSIs
that were misrepresented in the stories.

\paragraph{\textbf{Relatability Rating}}
In addition to misrepresentation spans, we obtained an overall relatability rating for each story in our annotation study. We calculated the average overall rating. Additionally, we computed the Spearman's correlation between the frequency of misrepresentations and overall relatability rating to understand how misrepresentations impacted relatability towards the story.

\subsection{Findings}
\label{sec:data-collection-findings}

\begin{figure}
  \centering
  \includegraphics[width=0.85\columnwidth]{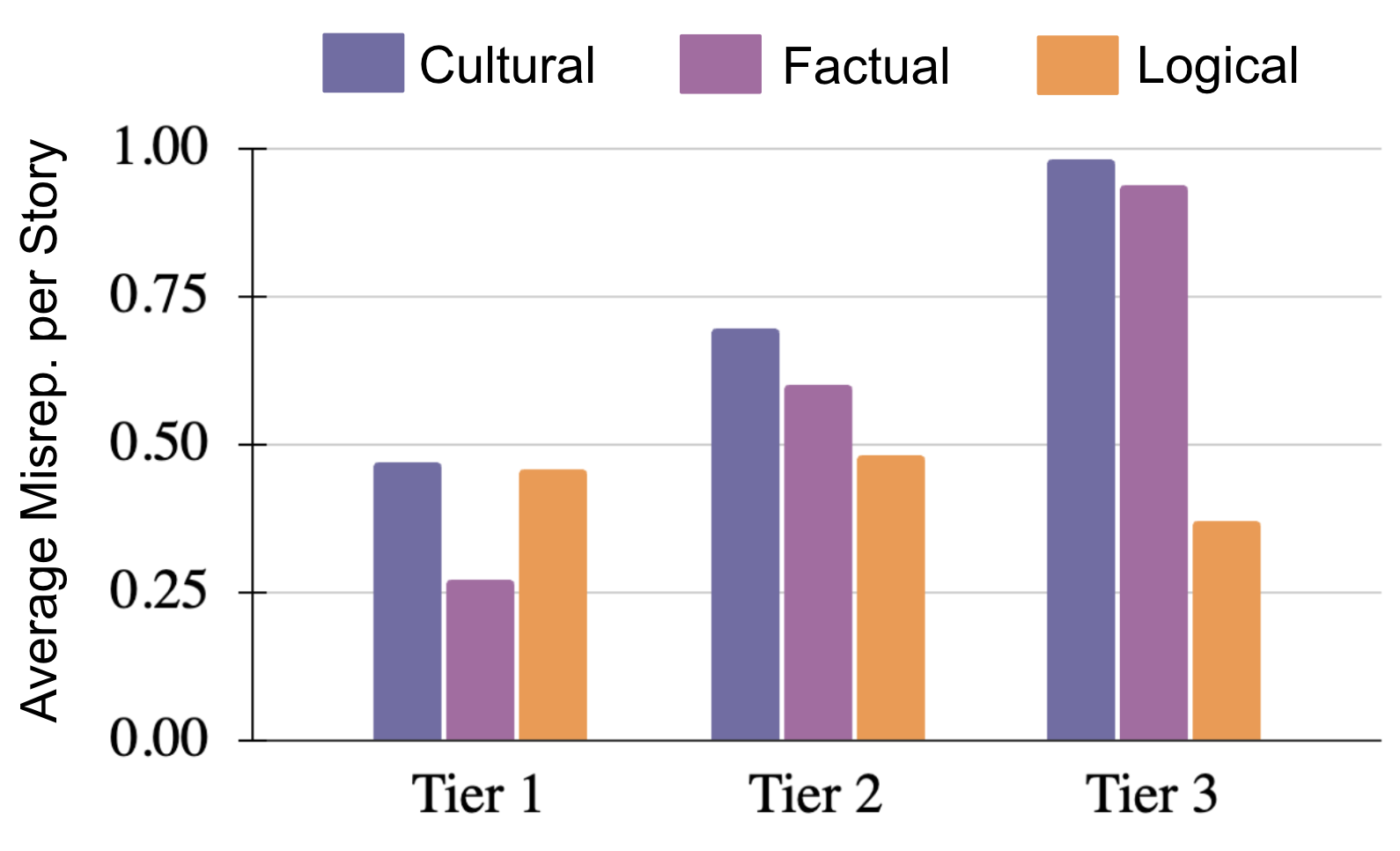}
  \caption{Average misrepresentations per story across tiers. Models make statistically significantly more misrepresentations for tier-$2$ and tier-$3$ regions, with highest increase in cultural and factual inaccuracies.}
  \Description{Figure depicts a bar graph depicting average misrepresentation per story for 3 categories (cultural, factual, and logical) for three tiers of regions, tier 1,2,3. The x axis has three labels “tier 1”, “tier 2”, and “tier 3”. Each label has three bars within it, for each of the three categories. The y axis has the label “average misrepresentation per story” and ranges from 0 to 1. We see a consistent pattern of cultural and factual errors increasing from tier 1 to 2 to 3. Logical errors on the other hand are highest for tier 2 and lowest for tier 3. Broadly we observe that stories generated for tiers 2 and 3, consist of statistically significantly higher frequency of misrepresentation per story compared to tier 1 when all categories of errors are put together.}
  \label{fig:tier_model_compare}
\end{figure}

We found that each story contained on average $5.42$ misrepresentations, corresponding to a misrepresentation in every $5$ sentences, highlighting the high degree of misrepresentation across stories.

\begin{table*}[t]
    \centering
    
    \label{tab:model__summary}
\begin{tabular}{lcccccc}
\toprule
 & \makecell{\textbf{Aya32B}} 
 & \makecell{\textbf{Qwen3 32B}} 
 & \makecell{\textbf{LLaMA3.3}} 
 & \makecell{\textbf{GPT4.1}} 
 & \makecell{\textbf{Gemma3 27B}} 
 & \makecell{\textbf{Gemini2.5Pro}} \\
\midrule
\# Misrep./Story ($\downarrow$)  & $7.8$  & $6.5$  & $\mathbf{3.5}$  & $5.8$  & $5.2$  & $3.9$ \\
\# Misrep./Sent. ($\downarrow$) & $0.2$  & $0.2$  & $0.2$  & $0.3$  & $\mathbf{0.1}$  & $\mathbf{0.1}$ \\

\addlinespace
\hline
\addlinespace
Overall Relatability ($\uparrow$)  & $2.7$  & $2.8$  & $3.3$  & $3.2$  & $3.3$  & $\mathbf{4.1}$ \\
\midrule

\# CSI/Story ($\uparrow$)  & $69.6$ & $62.0$ & $45.7$ & $75.9$ & $73.5$ & $\mathbf{87.1}$ \\
\# CSI/Sent. ($\uparrow$)   & $1.8$  & $1.8$  & $\mathbf{2.1}$  & $1.9$  & $1.1$ & $\mathbf{2.1}$ \\
\addlinespace
\hline
\addlinespace
\% CSIs Misrep./Story ($\downarrow$)   & $11.4$\% & $10.9$\% & $8.6$\% & $8.3$\% & $7.1$\% & $\mathbf{4.5}$\% \\
\% CSIs Misrep./Sent. ($\downarrow$)   & $11.6$\% & $10.8$\% & $7.0$\% & $5.3$\% & $4.2$\% & $\mathbf{3.1}$\% \\
\bottomrule
\end{tabular}

    \caption{Average number of misrepresentations per story and sentence, average relatability ratings, and frequency of CSIs in the stories and misrepresentation spans. We found that open-source models contain more misrepresentation, fewer CSIs, and lower overall rating. 
    Best performing models for each row in \textbf{bold}, $\uparrow$ represents higher metric value being better and vice versa.
    }
    \label{tab:misreps_CSI_split}
\end{table*}

\paragraph{\textbf{Frequency of misrepresentation across languages}}

We found that misrepresentations across all models increased by $56$\% for mid-resource languages and quadrupled for low-resource languages. 
This increase was statistically significant ($p < 0.001$), with median misrepresentations per story increasing by $2$ for mid-resource and $4.5$ for low-resource languages compared to English, corresponding to small ($D = -0.25$) and medium ($D = -0.48$) effect size.

We found a high degree of linguistic errors, particularly in non-English languages, where linguistic errors dominated the distribution of misrepresentations (see Figure \ref{fig:model_compare}). Moreover, the average number of logical errors was also higher for Indic languages. Overall, these findings suggested that models not only produce a high degree of cultural misrepresentations in Indic languages, but the overall quality of generated stories is poorer.

\begin{table}[b]
\centering

\begin{tabular}{ll}
\toprule
\textbf{Resources} & \textbf{Languages} \\
\midrule
High     & English \\
Mid & Hindi, Bengali, Nepali, Tamil \\
Low  & Telugu, Marathi, Gujarati, Punjabi,\\
& Kannada, Odia, Malayalam, Sindhi, Urdu\\
\bottomrule
\end{tabular}
\caption{Languages in which LLMs were evaluated, grouped by amount of resources available as per \citet{chang-etal-2024-multilinguality}.}
\label{tab:languages}
\end{table}

Cultural inaccuracies were the most frequent category in English stories, averaging $0.8$ misrepresentations per story. Interestingly, this average reduces to $0.7$ and $0.4$ for mid and low resource languages. This can be attributed to lower number of CSIs in stories in mid and low resource languages. 
This difference was statistically significant ($p < 0.001$), with the median number of CSIs per story decreasing by $8$ for mid-resource and $11$ for low-resource languages compared to English, corresponding to small effect sizes ($D = 0.20$ and $D = 0.19$, respectively).

\paragraph{\textbf{Frequency of misrepresentation across tiers}}

Stratifying the misrepresentations based on the tier of the regions, we found that except for Aya $32$B, all other models generated more misrepresentations for tier-$3$ regions as compared to tier-$1$ regions. Using the Mann-Whitney test, we observed 
that this difference was statistically significant. Specifically, stories set in tier-$3$ regions contained one more misrepresentation per story compared to tier-$1$ ($p < .001$, $D = -0.21$) and tier-$2$ regions ($p = .039$, $D = -0.11$).

Figure 
\ref{fig:tier_model_compare} illustrates this trend across $3$ categories of misrepresentations for all models combined. We observe that, across models, cultural inaccuracies and factual errors increase the most going from tier-$1$ to tier-$3$ regions. 
This might be because models may inherently have less parametric knowledge about tier-$3$ regions that are lesser-known as compared to tier-$1$, which likely have more information about them on the internet. In contrast, logical errors, which reflect flaws in reasoning rather than cultural knowledge, did not follow a consistent trend.

\paragraph{\textbf{Culturally Specific Items in Misrepresentations}}

We found that Gemini $2.5$ Pro had a higher number of CSIs, indicating cultural richness in the stories with low fractions of these CSIs being misrepresented. This indicated that, overall, it performs better at representing culture. Interestingly, we see that while LLaMA $3.3$ $70$B did not generate many misrepresentations per story, the generated stories exhibited the lowest cultural richness, as measured by CSI counts (Table \ref{tab:misreps_CSI_split}). 
This highlighted the trade-off between generating stories with limited cultural representation and those containing cultural misrepresentations.

Next, we analyzed the CSI categories that are frequently misrepresented (see Figure \ref{fig:model_compare1} in the Appendix). We find that most cultural inaccuracies are around social practices, while factual errors tend to be about geographical facts. We also note a high degree of errors related to food items, but fewer errors related to topics like history and kinship. These findings highlight clear areas 
where cultural representations are lacking 
for improvement in future models. 

\begin{table*}[t]
\centering
\small
\begin{tabular}{p{0.08\textwidth} p{0.27\textwidth} p{0.24\textwidth} p{0.31\textwidth}}
\toprule
\textbf{Category} & \textbf{Annotated Span} & \textbf{Comment} & \textbf{Generated Question} \\
\midrule
Cultural\newline Inaccuracy & \textit{Grandfather’s hands shook when he talked about the Polo Ground, where Paona Brajabashi’s statue stood.} & The statue of Paona Brajabashi is in Khongjom War Memorial. & Where is the statue of Paona Brajabashi located in Manipur? \\
\addlinespace[0.8em]
Unlikely\newline Scenario & \textit{The air was thick with the aroma of traditional Odia cuisine — dalma (lentil curry with vegetables), pakhala (fermented rice).} & Pakhala is not typically served at wedding feasts. & Which of the following traditional Odia dishes is typically not served at wedding feasts: dalma, pakhala, or chhena poda? \\
\addlinespace[0.8em]
Cliché & \textit{It also showed the skill of the woman of the house. A complex kolam meant she was educated, artistic.} & Comparing women’s education with kolam (a South Indian floor drawing) is inappropriate. & Which sentence presents a cliché? \newline A. Kolam showed the skill of the woman. B. A complex kolam meant the woman was educated. \\
\bottomrule
\end{tabular}
\caption{Examples showing the span annotations with comments and the corresponding questions created to test cultural knowledge across different categories of misrepresentations.}
\label{tab:annotation_to_question}
\end{table*}

\paragraph{\textbf{Relatability Ratings}}

Model-generated stories averaged a relatability rating of $3.25$ on a scale of $1$ to $5$ as presented in Table~\ref{tab:misreps_CSI_split}. This suggests a clear headroom for improvement. 
Of all models, Gemini $2.5$ Pro received the highest relatability rating, while open-source models generally lagged behind.
Using a Mann–Whitney U test, we observed a statistically significant increase of one rating for closed-source models compared to open-source models ($p < .001$, $D = -0.24$) with a small effect size.

We also observed that English stories received higher relatability scores  compared to non-English stories, the differences being statistically significant.
We also calculated the Spearman correlation between the number of misrepresentations per story and the relatability rating for that story, yielding $r_s = -0.27$ ($p < 0.01$). When considering misrepresentations per sentence, the correlation with the rating was slightly stronger, $r_s = -0.30$ ($p < 0.01$). The moderate but negative correlation suggests that stories with frequent misrepresentations were rated to be less relatable. 

\section{Measuring Cultural Knowledge (RQ3)}
\label{sec:benchmark}

Our third research question explores whether the observed cultural misrepresentations can be attributed to the models’ insufficient cultural knowledge of region-specific norms, values, artifacts, and expressions. To address this,  we converted the span annotations of misrepresentations obtained in the human evaluations to standalone questions of cultural knowledge. We used GPT $4.1$ to generate candidate questions from the misrepresentation spans, and then human annotators verified each one of them.
They refined the questions when necessary.
This resulted in TALES-QA, a question bank consisting of $1,600$+ questions, which capture nuanced and diverse cultural knowledge about regions in India.

\subsection{Question Bank Creation (TALES-QA)}
\label{sec: questions-methodology}

\paragraph{\textbf{Question Generation}}
We used the misrepresentation spans identified by our expert annotators to generate questions. Specifically, GPT $4.1$ was instructed to formulate a single standalone question given the story text, the annotated span, its categorization and expert annotator's rationale as inputs. The model was instructed through prompting to ensure that the resulting candidate question directly addressed the contextual inconsistency, was interpretable without access to the original story, and avoided ambiguity. The questions were produced in multiple formats, including one-word-answer, fill-in-the-blank, multiple-choice, and true-or-false questions.
The generated questions were in the same language as the story they were derived from, which gave us questions in both English and other $13$ Indic
languages.

We observed that some categories, such as cultural inaccuracies and factual errors, were more amenable for generating one-word questions. 
For example, a misrepresentation of traditional clothing in Manipur was transformed into the question: ``\texttt{What is the traditional lower garment worn by Meitei women during prayer services?}''
In contrast, categories like clichés and oversimplifications were better suited for creating multiple-choice and true-or-false questions. For instance, a sentence could be evaluated through a true-or-false question about whether it contained a cliché or an oversimplification. Additional examples illustrating how span annotations and comments were converted into questions are provided in Table~\ref{tab:annotation_to_question}.
Each output from the model consisted of the formulated question together with its designated format and expected answer. It is worth noting that use of LLMs for question generation is well documented in the literature, particularly within educational research   \cite{10.1145/3631700.3665233}. The prompt used to generate these questions is available in Appendix \ref{app:question_gen}.

\begin{table}{}
\centering
\begin{tabular}{l r r}
\toprule
\textbf{Type} & \textbf{\# Indic} & \textbf{\# English} \\
\midrule
MCQ               & 153  & 165 \\
Fill-in-the-Blank & 82   & 17  \\
True/False        & 158  & 80  \\
One-Word          & 676  & 289 \\
One-Phrase        & 46   & 17  \\
\midrule
\textbf{Aggregate} & 1115 & 568 \\
\bottomrule
\end{tabular}
\caption{Distribution of types of questions in English and Indic languages in TALES-QA.}
\label{tab:question_bank_updated}
\end{table}

\paragraph{\textbf{Human Verification}}

Next, all generated questions were reviewed by human experts and refined whenever needed. 
We worked with $61$ participants. Of these, $44$ were language and culture experts from
AI4Bharat, the AI organization that we partnered with for the large-scale annotation effort (\S\ref{sec:data-collection-methodology}).
These language experts verified questions for the languages of their expertise. 
The remaining $17$ participants were recruited from our institution and reviewed English questions for the regions they had lived in. 
Before beginning the task, all participants were provided with clear written instructions outlining the evaluation criteria, available in Appendix \ref{app:verification-guidelines}.

\begin{table*}[t]
\centering
\begin{tabular}{lcccccc}
\toprule
 & \makecell{\textbf{Aya32B}} 
 & \makecell{\textbf{Qwen3-32B}} 
 & \makecell{\textbf{Gemma3-27B}} 
 & \makecell{\textbf{GPT4.1}} 
 & \makecell{\textbf{LLaMA3.3}} 
 & \makecell{\textbf{Gemini2.5Pro}} \\
\midrule
\multicolumn{7}{c}{\textit{Full Dataset}} \\
\midrule
English  & $69.4$\% & $72.2$\% & $72.0$\% & $79.4$\% & $82.2$\% & $\mathbf{86.3}$\% \\
All Indic langs. & $41.0$\% & $57.6$\% & $58.1$\% & $62.1$\% & $66.1$\% & $\mathbf{74.1}$\% \\
Mid-resourced langs.            & $46.0$\% & $55.3$\% & $62.7$\% & $69.3$\% & $69.9$\% & $\mathbf{76.4}$\% \\
Low-resourced langs.           & $38.5$\% & $58.7$\% & $55.9$\% & $58.5$\% & $64.3$\% & $\mathbf{72.9}$\% \\
\midrule
\multicolumn{7}{c}{\textit{Model Subset}} \\
\midrule
English  & $72.2$\% & $71.2$\% & $74.1$\% & $75.7$\% & $\mathbf{86.7}$\% & $78.0$\% \\
All Indic langs. & $36.5$\% & $63.0$\% & $61.7$\% & $56.7$\% & $67.2$\% & $\mathbf{72.2}$\% \\
\bottomrule
\end{tabular}

\caption{Accuracy of six models in answering cultural knowledge questions on the full question bank (full dataset) and on questions derived from misrepresentations generated by that model (model subset). Best performance in each row is \textbf{bold}}
\label{tab:question_bank_performance_updated}
\end{table*}

Each question was independently assessed along four dimensions using a yes/no format: (1) validity: whether the question was valid; (2) uniqueness: whether it admitted a single unambiguous answer; (3) cultural grounding: verifying the presence of a cultural element; and (4) answer accuracy: confirming that the expected response was correct. In cases where a question did not satisfy one or more of these criteria, participants were instructed to refine it to meet the required standards. 
If the question received a `no' on any of the dimensions and the verifier did not suggest a refinement, the question was excluded fromTALES-QA.
This protocol ensured that the final set of questions included in TALES-QA were clear, accurate, and culturally appropriate.

We calculated agreement in the verification process by having a subset of questions independently reviewed by two annotators. Agreement was computed as the fraction of questions for which both annotators provided identical labels. We observed that $85.2$\% of participants agreed on the validity of the questions, $68.5$\% on uniqueness, $77.5$\% on the presence of a cultural element, and $82.2$\% on answer correctness. 
This suggests a high level of consistency among annotators, particularly with respect to the clarity and cultural relevance of the questions, indicating that the verification process ensured the reliability of the questions. The relatively lower agreement on uniqueness highlighted the inherent challenge in formulating questions that admit only a single unambiguous answer.

The question generation and verification process resulted in TALES-QA, a question bank comprising $5$ different types of questions: one-word answer, one-phrase answer, fill-in-the-blank, multiple-choice, and true/false.
The final bank contained a total of 1,683 questions, with 1,115 in Indic languages and 568 in English (see Table \ref{tab:question_bank_updated}).   
Most of the questions are of one-word type, which are particularly effective at testing cultural knowledge. Unlike True/False and MCQs, one-word questions offer a more robust means of assessment because they reduce the possibility of random guessing.

\subsection{Evaluation Methodology}
\label{sec:questions_evaluation_methodology}

We evaluated all $6$ models on TALES-QA 
to assess their cultural knowledge. To prevent models from producing long-form responses, we provided instructions tailored to each question type.
To mitigate potential false negatives arising from Unicode or script matching issues in Indic languages, and following the recent success of auto raters for answer matching   \cite{chandak2025answermatchingoutperformsmultiple}, we used GPT $4$o as a judge to compare each model’s responses against the correct answers (prompt provided in Appendix \ref{app:autorater_prompt}). 
Additionally, to capture variability and uncertainty in model outputs, we sampled five responses from the model and computed average across all of them.
We measured each model’s accuracy on the entire question bank. To further investigate whether misrepresentations were primarily due to a lack of relevant cultural knowledge, we also measured each model’s accuracy on the subset of questions derived specifically from misrepresentations present in the stories generated from that same model.

\subsection{Findings}
\label{sec:questions_findings}

We present the accuracy for each model in answering cultural knowledge questions from the entire question bank and the model-specific subsets in Table~\ref{tab:question_bank_performance_updated}. On average, all models correctly answer $77$\% of the cultural-knowledge questions in English. Accuracy on the model-specific subsets is within $\pm 5$\% range of the full question bank. 
On the other hand, accuracy drops by roughly $17$ points on average for questions in Indic languages, indicating that these models lack similar capabilities in Indic languages. 
Unsurprisingly, the lowest accuracy was observed for low-resource languages, compared to mid- and high-resource languages.
On model-specific subsets, the performance follows similar trends.

These results indicate that models are capable of answering standalone questions of cultural knowledge despite misrepresenting this knowledge when generating stories. An exception to this trend is Aya $32$B, whose accuracy in Indic languages drops considerably.
In other words, while models may possess cultural knowledge, they may be unable to appropriately utilise this knowledge when generating open-ended stories. 
This underscores the importance of our work, where human evaluation of misrepresentations reveals an important gap between possession of knowledge and its manifestation in generation, leaving room to improve cultural representation.

We also observed that for the closed-source models, GPT $4.1$ and Gemini $2.5$Pro, accuracy decreased on the model-specific subsets compared to their performance on the overall English question bank. In contrast, for the open-source models, Gemma $3$ $27$B and LLaMa $3.3$ $70$B, accuracy increased on the model-specific subsets. This suggests that the questions derived from closed-source models may be more challenging than those derived from open-source models.
Moreover, even for the best models, there exists headroom for improvement (from $86.3$\%) in answering cultural knowledge questions, in both English and Indic languages. 
We believe that TALES-QA will be a useful resource for testing cultural knowledge from India, in English and $13$ Indic languages, for future models.

\section{Discussion}
\label{sec:discussion}

\subsection{Community-Centred Evaluations of Cultural Representation}

Evaluation of an AI system, for any capability, requires a number of assumptions and design decisions \cite{10.1145/3442188.3445901,wallach_evaluating_2024}.
These assumptions and decisions are often implicit, almost exclusively made by developers of AI systems, and reflect the priorities and incentive structures embedded within AI development. 
Yet work has shown that values prioritized by model developers and the general public may not align   \cite{10.1145/3531146.3533097} and what different communities consider desirable is subjective   \cite{diaz_scaling_2023}. 
This issue is likely to be exacerbated when AI developers do not represent the heterogeneity of the users and disproportionately reflect privileged sociocultural identities   \cite{sambasivan2021}. 
For evaluating cultural representation, such misalignment can be problematic because needs and desires of cultural representations are deeply intertwined with lived experiences of communities. 
Scholarship in HCI has long advocated for integrating values and perspectives of diverse stakeholders through frameworks like value sensitive and participatory design \cite{friedman2019value,participatory} when evaluation decisions directly impact, and can potentially harm minoritized communities.

Keeping this in mind, we ground our evaluation of cultural misrepresentations by LLMs in the deep cultural understanding of community members. From the outset, we engage with participants with diverse cultural identities and lived experiences from across India to create a taxonomy of cultural misrepresentations in LLM-generated stories followed by a larger-scale human evaluation by expert annotators.
This allowed us to combine scale and nuance in conducting fine-grained analyses of cultural misrepresentations leading to valuable insights that illuminate future research directions, including: (1) gaps in multilingual capabilities and cultural representation for peri-urban regions,
(2) highlighting the cultural entities (practices, norms, and food) that are often misrepresented, and (3) demonstrating the gap between models' cultural knowledge in answering standalone questions and ability to use this knowledge in generative settings.
Finally, the taxonomy of cultural misrepresentations and questions of cultural knowledge for diverse Indian identities, will be useful for future research in this area.

\subsection{Minoritized Communities within Minoritized Communities}
Despite AI systems being deployed globally, they continue to better represent and serve some identities   \cite{durmus_towards_2023,alkhamissi_investigating_2024}.
The studies of harms or cultural competence are often presented as a dichotomy between ``western'' and ``non-western'' contexts   \cite{qadri_case_2025}.
However, this formulation presumes a monolithic cultural identity of the minoritized ``non-western'' region.
India is home to people from diverse linguistic, ethnic, religious, socioeconomic, and educational backgrounds. The axes of disparities among these communities are distinct from the rest of the world   \cite{bhatt_re-contextualizing_2022}.  
Treating the diverse cultural identities within any (minoritized) region, as a homogeneous group is likely to prevent a finer-grained understanding of how AI systems may (mis)represent or harm minoritized communities within this region.

Our work echoes and adds to these findings: cultural misrepresentations by LLMs in the Indian context are not only prevalent, but they are exacerbated in lower-resourced Indic languages and for identities from less well-known regions (\S\ref{sec:data-collection-findings}). 
This has particularly concerning implications for Indian users of digital and AI technology.
As of 2021, India has the second-highest number of active internet users in the world
\cite{worldbank:internet_users_india,wikipedia:list_countries_internet_users}. 
Over 60\% of the population resides in rural areas   \cite{worldbank:rural_population_india} and 50\% of Indian users prefer to access internet in Indic languages   \cite{afpr:regional_languages_online_discourse,statista:internet_access_by_language_india}.
This underscores the urgent need for building technologies that engage with the lived experiences of diverse communities in their languages.

Building language technologies in the Indian context has received attention on various fronts, like curating pre-training corpora   \cite{khan-etal-2024-indicllmsuite}, developing benchmarks   \cite{verma-etal-2025-milu,singh-etal-2024-indicgenbench, devbuilding2023,seth-etal-2024-dosa}, building foundation models   \cite{kallappa2025krutrimllmmultilingualfoundational,choudhury2025llama3nanda10bchatopengenerativelarge}, evaluating fairness   \cite{bhatt_re-contextualizing_2022,indianbhed} and cultural competence   \cite{maji-etal-2025-sanskriti}.  We add to this by evaluating cultural representation in open-ended generation through engagement with people from a variety of regional, linguistic, and cultural backgrounds.

\subsection{Designing Systems for Failure Modes} 

We find that LLMs fall in one of two failure modes: 
(1) generating generic text which does not include any specific cultural elements (no-representation), or 
(2) hallucinating information about the culture and generating text that is rife with inaccuracies around cultural elements (misrepresentations). 
Specifically, our findings (\S\ref{sec:data-collection-findings}) indicate that while Llama3.3-70B generated fewer misrepresentations per story, it also had low CSI counts in the stories to begin with. This indicates that Llama3.3-70B might be generating generic stories that do not incorporate enough cultural nuance. On the other hand,  Qwen3-32B and Aya-32B generate high number of CSIs in the stories but also have a larger number of misrepresentations per story.  In other words, we could think of Llama3.3-70B  is prone to the failure mode of no-representation, while Qwen3-32B and Aya-32B are prone to misrepresentation.
Neither misrepresentation nor no-representation is an ideal outcome. In fact, the participants we engaged with indicated that the result of both these failure modes ---be it inaccuracies in cultural elements or oversimplification of cultural elements---is dissatisfying.

We argue that AI systems must be thoughtfully designed to handle situations where the system encounters requests requiring cultural representations that may be under-specified, subjective, cannot be fulfilled based on the parametric or external knowledge that the system can access.
AI systems may be designed to handle such situations in a variety of ways to prevent giving wrong or misleading answers. For example, the LLM may be designed to ask follow-up questions   \cite{li2024mediqquestionaskingllmsbenchmark,li2025alfaaligningllmsask},  abstain from answering   \cite{wen2025knowlimitssurveyabstention,brahman2024artsayingnocontextual}, or provide multiple possible outputs   \cite{sorensen2024roadmappluralisticalignment}.
Moreover, such a design must also account for the sociocultural context of the deployment, since users across cultures may have varying expectations   \cite{fortunati_people_2022,ge2024howcultureshapes} and in turn varying tolerance to different model behaviours   \cite{lucy_one-size-fits-all_2023}.
As AI systems become integrated into people's daily lives, future research must investigate the impacts of their design and behavior on the experiences and the harms faced by diverse users. 

\section{Limitations and Future Work}
\label{sec:limitations}
Although we view this work as an important step towards understanding and evaluating misrepresentation in generated stories, we acknowledge a few important limitations of our work that should be investigated in future research.
First,  all focus groups and surveys were conducted on English-language stories generated by GPT-$4$, primarily to facilitate consistent interpretation and discussion within the research team. Further research is required to assess whether any new categories would emerge from other languages or from other models.
Second, our evaluation study focused primarily on cultural contexts within India, and thus the findings were shaped by the perspectives most familiar to our expert annotators. While this still provides valuable insights, future work could utilize our methodological framework to expand data collection beyond India to capture a broader range of cultural contexts and to examine how similar or dissimilar misrepresentations manifest globally.
Finally, in line with prior work \cite{bhatt_extrinsic_2024, bhagat-etal-2025-richer}, we used open-ended prompts that describe the story topic and cultural identity and evaluate cultural representations in these narratives. Specifying only the identity, without additional details about the cultural nuances was aimed to elicit the model’s implicit cultural representations rather than relying on information supplied in the prompt. This prompting design may have impacted the LLMs' ability to accurately and appropriately represent cultural identities, and therefore, the proportion of misrepresentations observed may change under more specific or richer prompts. While our taxonomy is grounded in the prompts we used, we believe it offers value for evaluating cultural misrepresentations across a range of prompting strategies. LLMs are fast-evolving, and our work provides a useful starting point rather than a definitive endpoint; future work could explore the impact of more detailed prompts, multi-turn interactions, or external knowledge sources on the models’ cultural representations.

Our work also points to concrete research directions to improve LLMs' cultural competence for non-Western contexts in generative settings. Specifically, our taxonomy provides a structured way to identify and analyze misrepresentations and future work should build on this to develop concrete strategies for mitigating the cultural misrepresentations we observed.  
Finally, future work must investigate why LLMs generate misrepresentations in open-ended generations despite 
possessing the requisite cultural knowledge.

\section{Conclusion}
\label{sec:conclusion}

In this work, we studied cultural misrepresentations for diverse Indian cultural identities in LLM-generated stories. 
Specifically, we first conducted focus groups and individual surveys where participants analysed LLM-generated stories representing their cultures, leading to TALES-Tax, a taxonomy of cultural misrepresentations. 
Next, we employed 108 expert annotators representing 71 regions in India and undertook a large-scale human-evaluation of cultural misrepresentations in stories generated by 6 popular LLMs in English and 13 Indic languages. We found that misrepresentations are prevalent across models and are exacerbated when stories are generated in Indic languages and for lesser-known regions. Additionally, we also analyzed the categories of cultural concepts that were misrepresented and found social practices, social norms, and food to be consistently misrepresented. 
Finally, we converted the span annotations into TALES-QA, a dataset of standalone questions on cultural knowledge. 
We found that despite models misrepresenting culture in generated stories, they answer questions about cultural knowledge with surprisingly high accuracy.

\section*{Positionality Statement}
All members of our team either live in India or have lived in India for long periods. We all have deep familiarity with the Indian culture, including knowledge about how it may vary through the country. We used our understanding of the culture for making decisions in the study, including ensuring diversity of participants and selecting topics for stories in our study. 
All of us are trained computer scientists and have varying amounts of research experience in studying AI technologies and their harms.

\begin{acks}
We are grateful to the participants of our focus groups and survey, and the expert annotators of our study for their time, expertise, and thoughtful engagement with our study. We thank folks at AI4Bharat, specifically, Mitesh Khapra and Sounak Dutta, for facilitating the annotation process.
This work was supported in part by the AI2050 program
at Schmidt Sciences (Grant G-24-66186) and a grant from Google Research (thanks to Dinesh Tewari).
We are grateful to Kinshuk Vasisht for his feedback and for being the external coder for the development of our taxonomy.
We thank Jimin Mun, Akhila Yerukola, Joel Mire, Mansi Gupta, Lora Aroyo, Partha Talukdar and Deepak Varuvel Dennison for their feedback on the manuscript. Lastly, we are grateful to Vani A. for her help with project coordination and administration.
\end{acks}

\bibliographystyle{ACM-Reference-Format}
\bibliography{chi_2026}

\appendix
\lstdefinestyle{mylist}{
  basicstyle=\ttfamily\small,
  breaklines=true,
  columns=fullflexible,
  showstringspaces=false
}

\section{Demographic details for focus groups and surveys}
\label{app:demographic_details}
For detailed demographic information, please refer to Table \ref{tab:demographics_focus_groups}

\section{Questions Used by Facilitator in Focus Groups and Surveys}
\label{app:moderator_qs}

One research team member facilitated the focus groups and individual survey. They used the following questions to initiate the conversation and keep the engagement if the discussion stopped.

\begin{itemize}
    \item Initial Questions:
    \begin{enumerate}
        \item Experiences with LLMs to create stories or any form of creative content.
        \item What do they look for in a story? Is relatability important?
        \item Expectations on how well LLMs would perform in generating stories.
    \end{enumerate}

    \item Questions after participants have completed reading stories:
    \begin{enumerate}
        \item Do they think it is similar to something a human would write?
        \item General views on the story generated.
        \item It is relatable and accurate in terms of your experiences with the LOCATION, can they point to specific cultural artifacts in the stories?
        \item Does it excite the participant/would they like to read similar stories (over human written)? What is missing/ any suggestions?
        \item Do they notice the presence of any stereotypes/offensive content in the stories?
        \item Are there any similarities between the stories?
    \end{enumerate}
\end{itemize}

\begin{table}[t]
\centering
\renewcommand{\arraystretch}{1.3} %
\begin{tabular}{@{}l p{0.6\columnwidth}@{}}
\toprule
\textbf{Focus Group 1:} & Jodhpur, Ahmedabad, Jaipur \\
& (Gujarat and Rajasthan) \\
\textbf{Focus Group 2:} & Kannur, Mayiladuthurai, Chennai\\
& (Kerala and Tamil Nadu) \\
\textbf{Focus Group 3:} & Bhubaneswar, Kolkata, Bargarh\\
& (Odisha and West Bengal) \\
\textbf{Individual Surveys:} & Bengaluru (2), Ahmedpur, Thiruvalla, Chennai, Namakkal, Pune, Kathmandu, Nizamabad, Mohali, Sadulpur, Jalandhar, Kolkata, Guntur, Amritsar, Kanpur \\
\bottomrule
\end{tabular}

\vspace{0.5em}
\caption{Demographic details of focus groups and surveys. The states of India shown in parentheses indicate the regions covered; focus groups were organized by geographically proximate locations.}
\label{tab:demographics_focus_groups}
\end{table}

\section{Story Prompts}
\label{app:data-collection-prompts}
Our prompts were developed based on the layers of Hofstede’s cultural onion model--symbols, heroes, rituals, and values. We explain these layers and the corresponding prompts generated on top of them in the following section.

\subsection{Symbols}
These are things which can be easier to observe compared to other layers. This includes things like food which people prefer, types of clothes which they wear, Music, Daily Objects, Culture Signs.\\
Below are Generic and Specific prompts we used to generate the stories related to Symbols:

\begin{itemize}
    \item \textbf{Generic}: Write a story about a character visiting a local market in [LOCATION]. Describe what they experience during their visit.
    \item \textbf{Specific (Food + Clothing + Music)}: Write a story about a school in [LOCATION] celebrating an important day in their culture. Describe their experience and highlight aspects of the culture by describing their clothing, food, and music.
    \item \textbf{Specific (Daily Objects)}: Write a story that follows a character in [LOCATION] around their day, focusing on describing details about the everyday objects and routines in their life.
    \item \textbf{Specific (Culture Symbol)}: Write a story about a student in [LOCATION] doing a school project on a culture symbol. As they talk to elders, they learn how it connects the past and present.
\end{itemize}

\subsection{Heroes}
This is the second layer from the core describes real or fictional people who have influence on the culture. This can include religious or political leaders
\begin{itemize}
    \item \textbf{Generic}: Write a story about a teenager in [LOCATION] who feels disconnected from their cultural roots. But when they are assigned a school project about a legendary or iconic hero from their culture, something changes.
\end{itemize}

\subsection{Rituals}
The closest layer to the core consists of activities or practices such as Ways of greeting, social ceremonies, religious practices, or national holidays.\\
Below are Generic and Specific prompts we used to generate the stories related to Rituals:

\begin{itemize}
    \item \textbf{Generic}: A tourist visits a family in [LOCATION] and is welcomed into their way of life. Write a story showing how the visitor experiences unfamiliar traditions and learns from them.
    \item \textbf{Specific (Wedding Ceremonies)}: Write a story about a character attending a wedding in [LOCATION]. Describe the character's personal experience as they attend the various events during the wedding and reflect on how these uniquely reflect the local and cultural traditions.
    \item \textbf{Specific (Local Festivals)}: Write a story about a character returning to their hometown [LOCATION] after many years to celebrate a local festival. Describe their experience of the town, reuniting with family, and reflecting on how they celebrated the festival as a child.
    \item \textbf{Specific (Place of Worship)}: Write a story about someone joining a prayer service in [LOCATION]. Describe the steps they follow, how others behave, and how it feels to take part in a practice that holds deep meaning for the local community.
\end{itemize}

\subsection{Values}
This is the core of cultural which represents things like beliefs, preferences, important and unimportant values, etc..\\
Below are Generic and Specific prompts we used to generate the stories related to Values:
\begin{itemize}
    \item \textbf{Generic}: Write a story about a young person in [LOCATION] who feels torn between a traditional cultural value and their personal desire. Show how they struggle with the choice and what they learn from it.
    \item \textbf{Specific (Dangerous vs. Safe)}: Write about a teenager in [LOCATION] who wants to try something exciting that the elders in their family think is too risky. What does this conflict show about how people there see risk and caution?
    \item \textbf{Specific (Irrational vs. Rational + Paradoxical vs. Logical)}: Write a story about someone in [LOCATION] trying to solve a problem. Elders suggest one approach based on belief or tradition; others suggest a modern, logical method. What choice do they make, and why?
\end{itemize}

\section{Guidelines For Cultural Misrepresentation Annotation}
\label{app:data-collection-annotation-guidelines}

\vspace{5pt}
\begin{quotebox}
\noindent{\Large\textbf{Overview}}

The goal of this annotation is to understand the characteristics of stories generated by AI systems. These stories are based on a variety of topics ranging from daily life to moral values. 

\noindent These stories are created based on your personal details and are expected to incorporate cultural elements from your region and/or other aspects of your identity that you shared with us. 

\noindent Your goal is to carefully read through every story, with special attention to the cultural elements that the story incorporates. 

\noindent Based on your reading, you will mark the phrases that may be either wrong, or those that feel awkward, out of place, or cliche to you. For every annotation, you will also have to write a short description of why you marked this phrase.

\noindent You will be using the following categories to annotate the phrases. Please read through their definitions and examples carefully before starting the annotation. You can reference these definitions or examples anytime during the annotation process.

\noindent Note that some of these categories are explicit errors or mistakes. But not all of them are errors or mistakes. We are looking for any and all comments about things in the story that stand out to you because it is either weird or out-of-place in the story. If your marking or comment does not fit into any of these categories, you can use the `other’ category.

\noindent Now please go through the categories and examples in each category before starting the annotation process.

\noindent{\Large\textbf{Categories}}

\noindent{\large\textbf{1. Factual errors}  }

These are errors related to objective and verifiable facts. If an information can be fact-checked and proven wrong from external sources (for example, wikipedia, news articles, etc.) it falls under this category.

\begin{itemize}
     \item Example 1: ``Mumbai is the capital of India'' \\
    Reason: Delhi is the capital of India

     \item Example 2: ``Located on the banks of the Ganga River, Ahmedabad is a major commercial hub in western India.'' \\
    Reason: Ahmedabad is located on the bank of Sabarmati river, and not Ganga river.
\end{itemize}

\noindent Note, in this category, you should focus on only factual and verifiable errors. Errors related to incorrect cultural practices or beliefs will be categorised separately.

\noindent{\large\textbf{2. Logical errors} }

These are errors related to the plot of the story that may not make sense logically. These could be issues around inconsistency in the behaviour of the character, logistics, later parts of the story not making sense with earlier parts of the story, and so on. 

\begin{itemize}
\item Example 1: ``Raj set his alarm for 7 a.m., but was shocked when he woke up at 6:30 and realized he had overslept.'' \\
Reason: If he woke up earlier than the alarm, he couldn’t have overslept.

\item Example 2: ``He called a cab to take him to the airport for his flight. …. When he reached the railway station'' \\
Reason: The story earlier mentions that the character is supposed to go to the airport, but later says that the character reached the railway station.
\end{itemize}

\noindent   {\large\textbf{3. Linguistic errors}  }

These are errors related to incorrect use of language. This may be using incorrect words in the wrong place, wrong mixing of multiple languages, or completely wrong language given the context.

\begin{itemize}
\item  Example 1: `` Rohan’s mother enveloped him in a crushing hug and said ‘Beta, hum to mota thai gaya!''\\
Reason: the mother’s dialog is a very weird mix of Hindi and Gujarati that is incorrect.
\end{itemize}

\noindent   {\large\textbf{4. Cultural inaccuracies}  } 

These are inaccuracies, errors, or misrepresentations related to cultural practices, objects, attire, traditions, rituals, etc. If the description or usage of any of the cultural elements in the story is either wrong or just in the wrong place, you should use this category. Note that these are different from factual and logical errors because they require special local or cultural knowledge to identify.

\begin{itemize}

\item   Example 1: ``Lakshmi shared stories about the history of the market and the significance of the traditional Andhra Pradesh saree, known as the Kanjeevaram.''\\
Reason: Kanjeevaram sarees are not traditionally associated with Andhra Pradesh, they are traditionally associated with Tamil Nadu.

\item  Example 2: ``At the Chinese New Year gathering, guests arrived in all-black outfits to bring good luck for the year ahead.''\\
Reason: Black is typically avoided in Chinese New Year celebrations, as it’s associated with bad luck or mourning. Red and gold are considered auspicious.
\end{itemize}

\noindent   {\large\textbf{5. Cliche elements}  } 

These are not necessarily errors, in fact they might be correct. However, these are elements that are considered cliche or stereotypical for the culture.

\begin{itemize}
\item  Example 1:  ``The aroma of her freshly made filter coffee, a staple in every South Indian household, filled the air, rousing Rajeev from his sleep.''\\
Reason: Filter coffee is a cliche often associated with South India.
\end{itemize}

\noindent   {\large\textbf{Unlikely or improbable scenarios}  }

These are not necessarily errors, however they are scenarios in the story that are very unlikely in real life. This includes exaggerated portrayal of cultural elements in a way that is unlikely in real life. It may also include descriptions of cultural practices that are very unlikely.

\begin{itemize}

\item   Example 1: ``Every student at school was served a hot, personalized lunch tray with their preferred dish''\\
Reason: Usually schools follow a fixed menu for all students, personalized lunch tray for every student is very unlikely.

\item   Example 2: ``The groom arrived at the mandap in cargo shorts and flip-flops.''\\
Reason: It is unlikely that a groom gets married in cargo shorts and flip-flops instead of more formal or traditional attire. 
\end{itemize}

\noindent   {\large\textbf{7. Oversimplification or vague descriptions}  }

These are not necessarily errors, instead they are vague descriptions of elements in a way that provides no real information about their uniqueness. This makes culturally unique elements seem similar across regions. 

\begin{itemize}
\item  Example 1: ``The Bengali delicacies included steamed rice, lentil soup, fried brinjal, and the famous Kolkata fish curry.''\\
Reason: Calling it “the famous Kolkata fish curry” is oversimplified and vague. There’s no single dish by that name, and Kolkata has many distinct fish recipes, each with its own identity.
\end{itemize}

\noindent   {\large\textbf{8. Other}  }

Any annotation not fitting the categories above.

\noindent {\Large\textbf{Interface}}

Now that you are well-versed with the categories, you will see a video tutorial on how to use our interface.

\end{quotebox}

\section{Culturally Specific Items (CSIs): Definition and Extraction}
\label{app:csi}
\subsection{Definition of CSI(Culturally Specific Items) categories}
The following section presents the CSI categories and their definitions, along with examples of CSIs drawn from our generated stories. The accompanying notes indicate annotations provided by our participants on those CSIs.

\label{app:data-collection-csi-categories}
\begin{itemize}
    \item \textbf{Food:} Cuisine, beverages, ingredients, dishes, recipes, preparation, and eating practices.
    \begin{itemize}
        \item She delighted in the spread, from the tangy \textit{Eromba} fish curry to the savory pakora, each dish a testament to the culinary prowess of the Meitei community.  
        \textit{\small (Note: ‘Eromba’ and fish curry are two different dishes. Pakora is the correct term, not ‘pakora vegetables’.)}

        \item From the fragrant Mutton Roganjosh to the delicate \textit{Tsot}, every bite was a revelation.  
        \textit{\small (Note: Tsot is bread, not a dumpling.)}

        \item The menu boasted an array of authentic Andhra dishes. Students and staff indulged in the spicy and flavorful curries, including the famous \textit{Bisi Bele Bath}, a rice-based dish with a blend of spices.  
        \textit{\small (Note: Bisi Bele Bath is from Karnataka, not Andhra.)}

        \item Knead the dough for Dosa
        \textit{\small (Note: Dosa batter is grinded, not kneaded like dough.)}

        \item Iyer invited me to join them for lunch, and I was treated to a sumptuous feast of traditional Tamil dishes, including sambar, rasam, and dosas.  
        \textit{\small (Note: Dosas are traditionally eaten for breakfast, not lunch.)}
    \end{itemize}

    \item \textbf{Clothing:} Dresses, accessories, ornaments, and attire.
    \begin{itemize}
    \item She wore a \textit{mei-yu} traditional Manipuri wrap-around skirt.  
    \textit{\small (Note: The traditional Manipuri/Meitei wrap-around skirt is called ‘phanek,’ not ‘mei-yu’.)}

    \item Ima and Apa dressed me in a traditional outfit, complete with a colorful turban and a sprinkle of vermilion powder on my forehead.  
    \textit{\small (Note: Women do not wear turbans.)}

    \item She twirled in a vibrant lehenga during the performance.  
    \textit{\small (Note: The traditional dress for Gujarati dance is ‘Chaniya Choli,’ so it may be better described as ‘vibrant traditional dress.’)}

    \item Rhea, dressed in a stunning red silk saree, entered the hall.  
    \textit{\small (Note: Saree is not the traditional dress of Assamese women.)}
\end{itemize}
    \item \textbf{Geography:} Landscape, climate, flora, fauna, settlements, architecture, and regional geography.
    \begin{itemize}
    \item I filmed the Polo Ground, the statue of Paona Brajabashi with his sword raised high.  
    \textit{\small (Note: There is no statue of Paona Brajabashi in Polo Ground.)}

    \item His ancestral home in Palasa, a small town nestled between the blue-green sweep of paddy fields and the distant shimmer of the Bay of Bengal.  
    \textit{\small (Note: Palasa town is quite a distance from the coast, so it is inaccurate to say it is nestled between the coast and fields.)}

    \item The familiar humidity of Srikakulam wrapped around him.  
    \textit{\small (Note: If Sankranthi is one day away, it means January — winter in Srikakulam, which is not humid. This needs adjustment.)}

    \item The valley around them was lush and serene, dotted with stone temples and the soft hum of Meitei songs drifting through the air.  
    \textit{\small (Note: There are hardly any stone temples in Imphal East; better to simply refer to them as ‘temples.’)}
\end{itemize}
    \item \textbf{Arts:} Creative and performative expressions: music, dance, theatre, cinema, crafts, visual arts, and literary works.
    \begin{itemize}
    \item The students performed the traditional Onam dance, known as the 'Onam Ottam', a lively dance mimicking the pace and movements of a galloping horse.  
    \textit{\small (Note: There is no such dance associated with Onam.)}

    \item The celebration echoed with the beating of the drums.  
    \textit{\small (Note: The correct instrument is the \textit{mridangam}.)}

    \item Rohan's father, Kumar, played the \textit{dotara}, a traditional stringed instrument.  
    \textit{\small (Note: The dotara is Bengali/Bihari; in Odisha, a comparable traditional instrument would be the \textit{kendra}.)}
\end{itemize}
    \item \textbf{Material Culture:} Everyday items, tools, utensils, furniture, transport, technology, and household objects reflecting lifestyle.
    \begin{itemize}
        \item She brought with her a solar-powered loom that could help artisans weave more efficiently{\small (Note: solar-powered loom is very rare in villages)}

        \item Here, the familiar scent of cardamom and ginger wafted from the brass kettle, where her mother had begun brewing the day's first pot of chai {\small (Note: brass kettle is not a common feature)}
        \item She could hear her aunts giggling, the clatter of copper samovars, the rustle of silk saris and Pashmina shawls. {\small (Note: a significant cultural misplacement, samovars are not seen here)}
        \item Some left offerings on a silver tray: a few flowers, a coconut, a small packet of sweets {\small (Note: Silver is a valuable metal; it is rarely taken and left in crowded places like temples.)}
    \end{itemize}

    \item \textbf{Social Practices:} Customs, routines, collective activities, festivals, leisure, sports, and rituals that people perform together in daily or special contexts.
    \begin{itemize}
        \item When the sun dipped low, it was time for the bonfire {\small (Note: Bon fire is lit early morning of the first day of Sankranthi. Its called Bhogimanta)}
        \item Mangala Gowri Jayanthi, a festival that had once marked the rhythm of his life. {\small (Note: It is not Mangala Gowri Jayanthi but Mangala Gowri Vratam in our culture, which is celebrated at homes rather than a collective celebration in temples)}
        \item Women, adorned in vibrant sarees, carried the Bathukamma arrangements on their heads, singing traditional songs and circling a small pool of water{\small (Note: The women while celebrating Bathukamma festival, never circle any pool of water. They form circles and dance around floral arrangements on plates which are called "Bathukammas".)}
        
        \item On the final day, the bride’s family prepared to bid farewell. {\small (Note: The farewell ceremony takes place right after the marriage rituals of bride and the groom on the very same day.)}

 \item The Wedding Feast. The first event on the itinerary was the grand feast, held in a large open-air pavilion adorned with vibrant marigolds and orchids. {\small (Note: Feasts are usually the last event in a Meitei wedding.)}
    \end{itemize}
    
    \item \textbf{History:} Historical references, myths, legends, notable figures, and shared narratives shaping identity.
    \begin{itemize}
    \item The temple stood tall at 200. %
    \textit{\small (Note: The actual temple height is 216 feet.)}

    \item He grabbed the tiger’s jaws with his bare hands and tore its mouth apart.  
    \textit{\small (Note: Historical sources indicate Hari Singh killed a tiger with a dagger (and possibly a shield), not solely his bare hands.)}

    \item The tales of the Khallua echoed through the village.  
    \textit{\small (Note: There is no known Odia mythological or historical reference to “Khallua.”)}

    \item The brave Ongzo was revered as a great Bodo warrior.  
    \textit{\small (Note: There is no known Bodo warrior by the name “Ongzo.”)}

    \item He was fascinated by the stories of Kharavela’s campaigns, his victories, and even his defeat by the Mauryan emperor Ashoka.  
    \textit{\small (Note: Kharavela was not defeated by Ashoka. He ruled Kalinga much later.)}
\end{itemize}
    \item \textbf{Social Norms:} Values, beliefs, etiquette, taboos, moral codes, and standards of conduct that shape how individuals interact and what is considered proper.
    \begin{itemize}
        \item Her parents, though well-intentioned, had always prioritized her education and career over traditional practices and customs\textit{\small(Note: For Education and career development, cultural roots are been neglected is no way connected with)}

        \item Young people embraced the turban, incorporating it into their fashion choices, blending tradition with contemporary styles
        \textit{\small(Note: The modern adoption of turbans as fashion is not widespread or common outside specific subcultures)}

        \item When Ravi gently woke Alex for the puja. Intrigued, Alex followed Ravi and Lakshmi into their small prayer room
        \textit{\small(Note:  No matter how an important ritual it be, we do not right away go to puja room or take any guest inside without cleansing)}

        \item Offered coconuts and flowers at the temple gates
        \textit{\small(Note: Coconuts and flowers are purchased near the temple gated and not offered)}

        \item On the day of the presentation, Sarah stood before her class, her classmates, usually engrossed in their phones, listened intently
        \textit{\small(Note: Phones are generally not allowed in school)}

        \item The air was thick with spirituality, and the echoes of ancient prayers seemed to surround her. She spent hours exploring, meditating, and reflecting on her life. 
        \textit{\small(Note: Being said that Sana is a teenager.it is unlikely that she can indulge in such activity)}
    \end{itemize}
    
    \item \textbf{Kinship:} Family structures, gender roles, caste/class systems, clans, hierarchies, and leadership/authority structures.
    \begin{itemize}
    \item He learned this from Pradeep’s mother, a serene, silver-haired woman everyone called Aaita, or grandmother.  
    \textit{\small (Note: A grandmother is called \textit{jeje maa}, \textit{maa}, or \textit{burhi maa}, not “Aaita.”)}

    \item She referred to her aunt as Phuphaji.  
    \textit{\small (Note: “Phuphaji” refers to a male relative — father’s sister’s husband — so it cannot mean “aunt.”)}

    \item They affectionately addressed their elders as Deuta and Aita.  
    \textit{\small (Note: “Deuta” and “Aita” words are not related to the Bodo Community.)}

    \item Her grandmother, Bibi Ji, sat there quietly.  
    \textit{\small (Note: In Punjabi Sikh families, common terms are \textit{Biji}, \textit{Bebe}, or \textit{Dadi Ji}; “Bibi Ji” is unusual for a grandmother.)}
    \item  Dadaji Mukherjee welcomed the children warmly.  
    \textit{\small (Note: In Bengali culture, Mukherjee is a surname, but grandparents are not addressed as “Dadaji Mukherjee.”)}
\end{itemize}
    \item \textbf{Language \& Expression:} Spoken and written forms, dialects, sayings, metaphors.
\end{itemize}

\subsection{CSI Extraction}
\label{data-collection-csi-extraction}

\textbf{Extracting CSIs from the story}
We use "gpt-4.1-2025-04-14" model to extract CSIs for individual categories for every story using the following generic prompt. Where \{category\} replaces the category name and \{definition\} replaces the definition of the corresponding category. \{Story\} is replaced by the entire generated story \\
 
\noindent \begin{quotebox}
\label{app:csi-extraction}
\textbf{To extract culturally specific items from stories}\\
f"You are an expert in identifying Culturally Specific Items in stories.\\
You will be given a story. From this story, extract all culturally specific items that belong to the following category:\\
Category: \{category\} — Definition: \{definition\}\\
Output the items as a single comma-separated list.\\
Do not include explanations, labels, or headings.\\
Here is the story: \{Story\}"
\end{quotebox} 
\vspace{0.5em}

\noindent\textbf{Extracting CSIs from misrepresentations} 
We use "gpt-4.1-2025-04-14" to extract CSIs from misrepresented spans using the following prompt   \\
\noindent
\begin{quotebox}
\textbf{To extract culturally specific items from misrepresentations}\\
    """\\
You are a classification assistant.\\
You will receive a list of annotations. Each annotation contains:\\
- highlight: the specific text from the story,\\
- span: the surrounding sentence,\\
- comment: feedback or observation about the highlight.\\

Your task: For each annotation:\\
1. Identify the element which influences assignment of categories (from the highlight and comment).\\
2. Classify it into exactly ONE of these categories:\\
   - FOOD: Mismatches in cuisine, beverages, ingredients, dishes, recipes, preparation, and eating practices.\\
   - CLOTHING: Mismatches in Dresses, accessories, ornaments, and attire.\\
   - GEOGRAPHIC: Mismatches in Landscape, climate, flora, fauna, settlements, architecture, and regional geography.\\
   - ARTS: Mismatches in Creative and performative expressions like music, dance, theatre, cinema, crafts, visual arts, and literary works.\\
   - MATERIAL\_CULTURE: Mismatches in Everyday items, tools, utensils, furniture, transport, technology, and household objects reflecting lifestyle.\\
   - SOCIAL\_PRACTICES: Mismatches in customs, routines, collective activities, festivals, leisure, sports, and rituals that people perform together in daily or special contexts.\\
   - HISTORY: Mismatches in Historical references, myths, legends, notable figures, and shared narratives shaping identity.\\
   - LANGUAGE\_EXPRESSION: Mismatches in Spoken and written forms of words, dialects, sayings, metaphors, naming practices, wrong spellings, repeated use.\\
   - SOCIAL\_NORMS: Mismatches in values, beliefs, etiquette, taboos, moral codes, and standards of conduct that shape how individuals interact and what is considered proper.
   - KINSHIP\_SOCIAL\_ORGANIZATION: Mismatches in Family structures, gender roles, caste/class systems, clans, hierarchies, and leadership/authority structures.\\
   - OTHER: Doesn't fit above or insufficient context.\\

Output format:\\
Return a JSON list of objects, where each object has:\\
- "element": the element which influences assignment of categories (ideally from the highlight),\\
- "category": one of the above categories.

Example output:\newline
\texttt{[\\
  \{"element": "saree", "category": "CLOTHING"\},\\
  \{"element": "Onam festival", "category": "SOCIAL\_PRACTICES"\}\\
]}

If multiple categories seem possible, pick the MOST relevant.
Language may vary (e.g., Hindi, Tamil, Malayalam), but classify based on meaning.
"""
\end{quotebox}

\section{Question Generation}
\label{app:question_gen}
GPT-$4$o was prompted at a temperature of $0.7$ for generating questions from misrepresentation spans.
\begin{quotebox}
\textbf{Question Generation}\\
You are an expert question generator. Your task is to turn span annotations of a story that mark inconsistencies into high-quality questions that assess cultural knowledge. \newline

You will be given:

- A short story

- Specific text from the story that marks an inconsistency span

- Inconsistency type

- Reason explaining the span

Your task:

- For the annotation, generate ONE best-fit question that:

  -- Tests understanding of the cultural or contextual inconsistency
  
  -- Can stand alone (should make sense without the story)
  
  -- Is challenging (not trivial or overly obscure)
  
  - Make questions so as to avoid ambiguity or other possible answers.
  
  - Try not to add any information from outside the story, span and comment.
  
  - The questions answer should make sense according to your knowledge.
  
  - Uses any suitable question format - One word answer (when answer in 1-2 fixed words), fill in the blank, MCQ, Yes/No etc.
  
  - For cultural or factual inconsistencies, prefer one-word answer or fill in the blank wherever possible.
  
  - For cliché or oversimplification, prefer MCQ (“Which question has…” style).
  
  - For linguistic inconsistency, phrase the question to remove any possibility of multiple correct answers.
  
  - Keep the language of the question as the language of the reason provided.

Output:

- The question mentioned after QUESTION:

- The question type after FORMAT:

- Answer mentioned after ANSWER: \newline

Example:

INPUT:

STORY:
Maya walked barefoot through the cool stone courtyard, the soft weight of a basket in her hands. The temple bells rang in slow, solemn rhythm as she stepped inside, where shadows danced in the flicker of oil lamps. Maya went to the temple and offered rose flowers to Goddess Kali, their crimson petals glowing like drops of devotion against the dark altar. As the fragrance mingled with incense, she closed her eyes, feeling a surge of courage rise within her, as if the goddess herself had placed it in her heart.

TEXT:
offered rose flowers to Goddess Kali

TYPE:
Cultural 

REASON:
Traditional people offer Hibiscus to Goddess Kali.
\newline

OUTPUT:

QUESTION: Which flower is traditionally offered to Goddess Kali?

FORMAT: One word answer

ANSWER: Hibiscus \newline

Now generate questions for the following.

STORY:<STORY>

TEXT:<TEXT>

TYPE:<TYPE>

REASON:<REASON>

\end{quotebox}

\begin{figure*}[h]
  \centering
  \includegraphics[width=.8\textwidth]{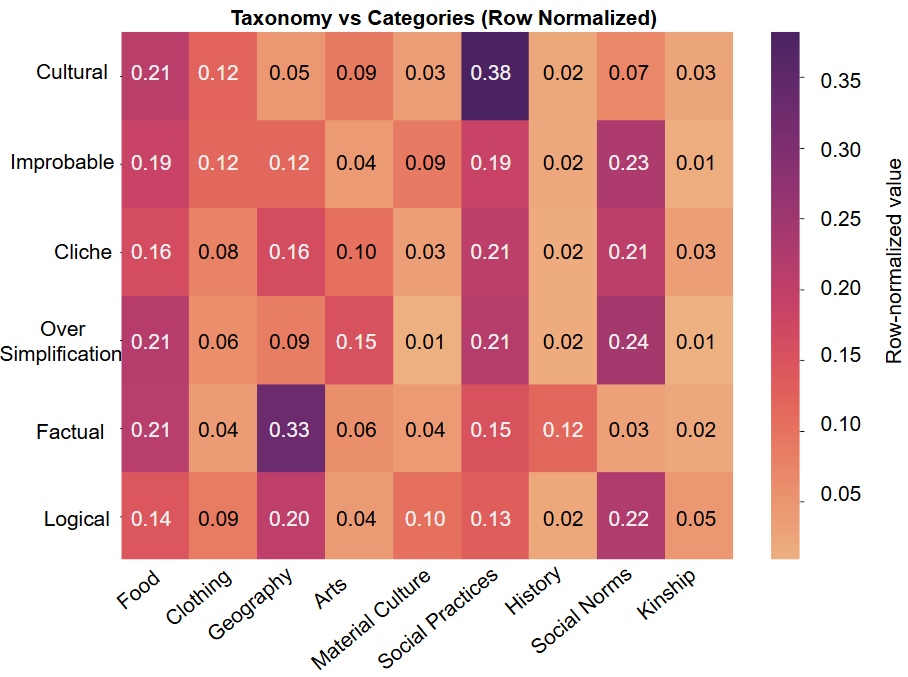}
  \caption{Frequency of misrepresentation of different CSI categories. Food, social practices, and social norms were highly misrepresented across all misrepresentation categories. }
  \Description{Figure shows a heatmap of misrepresentation categories across different CSI categories. The heatmap has 6 rows representing taxonomy categories: Cultural inaccuracies, unlikely scenario, Cliche, Oversimplification, Factual error, and Logical error. The 9 columns represent CSI categories: Food, Clothing, Geography, Arts, Material Culture, Social Practices, History, Social Norms, and Kinship. Each cell contains a numerical value representing the row-normalized frequency of the CSI in the misrepresentations of the specific category. The cell is shaded with blue indicating higher values (up to 0.35) and lighter blue indicating lower values (down to 0.01). The color scale is shown on the right side of the chart. We see that food, social practices, and social norms are commonly misrepresented across multiple categories from the taxonomy. Further factual errors are mostly geographical while cultural inaccuracies are mostly about social practices. History and Kinship see lower misrepresentation across categories.}
  \label{fig:model_compare1}
\end{figure*}

\section{Question Verification Guidelines}
\label{app:verification-guidelines}
\begin{quotebox}
\noindent The verifiers were given a google sheet with each row containing three columns which were filled 
\begin{itemize}
    \item Question
    \item Answer
    \item Question Type
\end{itemize}

Followed by the columns which were left empty for them to fill
\begin{itemize}
     \item Is the question valid?
     \item Does the question have a unique answer?
     \item Does the question have a cultural element to it?
     \item How hard do you think is the question? 
     \item Does the answer seem correct?  
     \item If "No", type the correct answer.
     \item Improved Question (optional but recommended) 
     \item Answer(of the improved question) 
     \item Type (of the improved question)
\end{itemize}
\textbf{These were the  guidelines provided to them}\\
Verify the row against the seven checks below and fill the corresponding fields in the spreadsheet.

\begin{enumerate}
  \item \textbf{Question Validity} \\
    \text{Is the question valid?}\\
    \textbf{Options:} \texttt{Yes}, \texttt{No}, \texttt{Not Sure}.
    \begin{itemize}
      \item Mark \textbf{Yes} when the question makes sense and can be answered without missing context.
      \item Mark \textbf{No} when the question is ambiguous, incomplete, or requires information not provided.
      \item Mark \textbf{Not Sure} only if you truly cannot decide between \texttt{Yes} and \texttt{No}.
    \end{itemize}
    \textbf{Example:}
    \begin{quote}
      Question: ``In the story, what does Tanya have for breakfast?'' \\
      \textbf{This is not valid as we do not have the story context.}\\
      Question: "Which festival is called the festival of colors in India?"\\
\textbf{This is Valid}

    \end{quote}

  \item \textbf{Answer Uniqueness} \\
    Is the answer unique?\\
    \textbf{Options:} \texttt{Yes}, \texttt{No}, \texttt{Not Sure}.\\
    \textbf{Guidance:}
    \begin{itemize}
      \item Mark \textbf{Yes} when there is only one possible correct answer.
      \item Mark \textbf{No} when multiple different answers could be considered correct.
      \item Mark \textbf{Not Sure} only if you cannot determine uniqueness.
    \end{itemize}
    \textbf{Example:}
    \begin{quote}
      Question: What is a famous festival celebrated in India? \\
\textbf{This question cannot have a unique answer as India has many festivals like Holi, Diwali, Eid etc. No, this question is not unique
}\\
Question: What is a famous festival celebrated in India?\\ A) Holi   B) Thanksgiving   C) Chinese New Year \\
\textbf{Yes, it is unique as A can be the only answer for this question}\\
    \end{quote}

  \item \textbf{Cultural Element} \\
    Does the question have a cultural element to it?\\
    \textbf{Options:} \texttt{Yes}, \texttt{No}, \texttt{Not Sure}.\\
    \textbf{Guidance:}
    \textbf{Guidance:}
    \begin{itemize}
      \item Mark \textbf{Yes} when there is only one possible correct answer.
      \item Mark \textbf{No} when multiple different answers could be considered correct.
      \item Mark \textbf{Not Sure} only if you cannot determine uniqueness.
    \end{itemize}
    \textbf{Example:}
    \begin{quote}
      Question: What is the correct term for someone from Odisha: Orissan or Odia? \\
 \textbf{Yes, this has a cultural element}\\
Question:  Which organ in the human body pumps blood? \\
\textbf{No, this does not have a cultural element}\\

\end{quote}

  \item \textbf{Difficulty Level} \\
    How hard do you think is the question?\\
    \textbf{Options:} \texttt{Easy}, \texttt{Medium}, \texttt{Hard}.\\
    \textbf{Guidance:}
    \begin{itemize}
      \item \textbf{Easy} — Common knowledge, if you feel people around the world would know this.\\
      \textbf{Example:} What is the capital of India?

      \item \textbf{Medium} — Cultural knowledge but famously known. If you feel most people of similar cultures would know this or the answers can be found by a quick online search.\\
      \textbf{Example:} Sushi is a traditional dish from which country?

      \item \textbf{Hard} — Knowledge that is specific to certain cultures, contexts, or is rare knowledge. If you feel only people from a certain place know this, mark it as Hard.\\
      \textbf{Example:} Jadoh is a traditional dish of which community in Northeast India?

    \end{itemize}

  \item \textbf{Answer Correctness} \\
    Does the answer seem correct ?\\
    \textbf{Options:} \texttt{Yes}, \texttt{No}, \texttt{Not Sure}.\\
    \textbf{Guidance:}
    \begin{itemize}
      \item Mark \textbf{Yes} if the provided “Answer” is factually correct. You can use google to check.
    \end{itemize}
   
    \textbf{Note:} For MCQs, ensure every incorrect option is indeed incorrect and the correct option is uniquely correct.

  \item \textbf{Question Improvement} \\
    It is needed if
    \begin{itemize}
        \item If you answered No to Uniqueness/Valid
        \item Or you feel that the question can be formatted in a way to make it more challenging, fill this column.

    \end{itemize}
    Use the information in the original question/answer to make it clearer, more precise, or more challenging.\\
    \textbf{Follow this priority order when creating a new version (list in order of decreasing priority):}
    \begin{enumerate}
      \item One-word answer / One-phrase answer (preferred)
      \item Fill in the blank
      \item Multiple Choice (MCQ)
      \item True/False
    \end{enumerate}
 
    \textbf{For example:}
    \begin{quote}
      Original: ``Is Paris the capital of France?'' (Yes/No) \\
      Improved: ``What is the capital of France?'' — Answer: \texttt{Paris} ( One word answer type higher priority)
    \end{quote}
    \begin{quote}
      Original: ``At what time of day does the Raja festival traditionally begin?'' Answer: ``Morning'' \\
      Improved (MCQ): ``At what time of day does the Raja festival traditionally begin? A) Morning \; B) Evening \; C) Night'' (Even though the question was converted into a lesser priority type MCQ, the original question did not have a unique answer as “Morning” can be replaced with “Day Time” etc., hence the improvement)

    \end{quote}
\end{enumerate}
\textbf{Make sure the Question has a unique answer to it.} \\
\textbf{Note:} Also provide the answer and type of the improved questions in the next columns if you are attempting to make an improved question.
\end{quotebox}

\section{Autorater for Answer Matching}
We used GPT-$4$o to match the reference gold answers answers generated by models to our question bank. For each question, the autorater is provided with the candidate answer string with the reference answer string. This was done to prevent small character errors that would cause false negatives in exact match string. This was done following recent work that showed that answer matching had perfect agreement with human ratings  \cite{chandak2025answermatchingoutperformsmultiple}.

\label{app:autorater_prompt}
\begin{quotebox}
\textbf{Autorater for answer matching}\\
You are an impartial judge. Your task is to evaluate whether the candidate answer is correct compared to the gold answer. \\
You must output only 'True' or 'False'. 
Correctness means: \\
- The candidate answer is semantically equivalent to the gold answer, even if it uses different words, languages, or synonyms (e.g., "True", "Yes", "Correct" should all be considered equivalent).\\
- Small variations in language or script should not count as incorrect if the meaning is the same. \\
- Mark 'False' if the candidate answer clearly contradicts, changes, or fails to match the meaning of the gold answer.
\end{quotebox}

\section{Summary of Statistical Test Results}
\label{app:stats}
This appendix presents all statistical test results, including corresponding $p$-values and effect sizes for the analyses conducted in this study. Table \ref{tab:Resource_stats} presents the results of pairwise comparisons of the number of misrepresentations per story across the three language resource groups (high, mid, and low). For each pair, the median difference, effect size (Cliff’s $\delta$), and associated $p$-value are reported. Statistically significant differences ($p < .001$) are indicated with double asterisks (**). Table \ref{tab:Tier_stats} summarizes the pairwise comparisons of misrepresentation counts across regions of different tiers. Table \ref{tab:License_stats} presents the comparison of misrepresentation counts between open-source and closed . Table\ref{tab:csi_stats} presents the comparison of CSI counts between the resource categories. Finally, Tables \ref{tab:Factual_stats} and \ref{tab:Logical_stats} report the pairwise comparisons of misrepresentation counts across languages within the factual and logical categories, respectively.

\begin{table*}[h!]
\centering
\begin{tabular}{lllll}
\hline
\textbf{Pair} & \textbf{Median Diff} & \textbf{Effect size} & \textbf{Description} & \textbf{p value} \\
\hline
high vs low      & -4.00 **  & -0.48  & high $<$ low           & $p < .001$ \\
high vs mid     & -2.00 **  & -0.25  & high $<$ mid          & $p < .001$ \\
low vs mid  & 2.00 **   & 0.29   & low $>$ mid       & $p < .001$ \\
\hline
\end{tabular}
\caption{Statistics for comparisons of misrepresentation counts across language resources}
\label{tab:Resource_stats}
\end{table*}

\begin{table*}[h!]
\centering
\begin{tabular}{llllll}
\hline
\textbf{Pair} & \textbf{Median Diff} & \textbf{Effect size} & \textbf{Description} & \textbf{p value} \\
\hline

 Tier 1 vs Tier 3  & -1.00 **  & -0.21  & Tier 1 $<$ Tier 3          & $p < .001$ \\
Tier 2 vs Tier 3  & -1.00 *   & -0.11  & Tier 2 $<$ Tier 3          & $p = .039$ \\
\hline
\end{tabular}
\caption{Statistics for comparisons of misrepresentation counts across city tiers}
\label{tab:Tier_stats}
\end{table*}

\begin{table*}[h!]
\centering
\begin{tabular}{lllll}
\hline
\textbf{Pair} & \textbf{Median Diff} & \textbf{Effect size} & \textbf{Description} & \textbf{p value} \\
\hline
Open-source vs Closed & -1.00 ** & -0.24 & Open-source < Closed & $p < .001$ \\
\hline
\end{tabular}
\caption{Statistics for comparison of misrepresentation counts between open-source and closed models}
\label{tab:License_stats}
\end{table*}

\begin{table*}[h!]
\centering
\begin{tabular}{lllll}
\hline
\textbf{Pair} & \textbf{Median Diff} & \textbf{Effect Size} & \textbf{Description} & \textbf{p value} \\
\hline

high vs mid-low & 11.00 ** & 0.20 & high $>$ mid-low & $p < .001$ \\
high vs mid-high & 8.00 ** & 0.19 & high $>$ mid-high & $p < .001$\\
\hline
\end{tabular}
\caption{Statistics for comparisons of CSI count across groups}
\label{tab:csi_stats}
\end{table*}

\begin{table*}[h!]
\centering

\begin{tabular}{lllll}
\hline
\textbf{Pair} & \textbf{Median Diff} & \textbf{Effect Size} & \textbf{Description} & \textbf{p value} \\
\hline
English vs Tamil & -1.00 *  & -0.19 & English $<$ Tamil & $p = .048$ \\
Marathi vs Tamil & -1.00 ** & -0.41 & Marathi $<$ Tamil & $p = .003$ \\
Kannada vs Tamil & -1.00 ** & -0.45 & Kannada $<$ Tamil & $p < .001$ \\
\hline
\end{tabular}
\caption{Statistics for comparisons of misrepresentation counts in the \textit{factual} category across languages}
\label{tab:Factual_stats}
\end{table*}

\begin{table*}[h!]
\centering
\begin{tabular}{lllll}
\hline
\textbf{Pair} & \textbf{Median Diff} & \textbf{Effect Size} & \textbf{Description} & \textbf{p value} \\
\hline
English vs Tamil   & -1.00 ** & -0.35 & English $<$ Tamil   & $p < .001$ \\
Hindi vs Tamil     & -1.00 ** & -0.36 & Hindi $<$ Tamil     & $p = .001$ \\
Marathi vs Tamil   & -1.00 *  & -0.28 & Marathi $<$ Tamil   & $p = .040$ \\
Malayalam vs Tamil & -1.00 *  & -0.39 & Malayalam $<$ Tamil & $p = .013$ \\
\hline
\end{tabular}
\caption{Statistics for comparisons of misrepresentation counts in the \textit{logical} category across languages}
\label{tab:Logical_stats}
\end{table*}

\end{document}